\documentclass[%
 reprint,
nofootinbib,
 amsmath,amssymb,
 aps,
]{revtex4-2}

\usepackage{graphicx}
\usepackage{dcolumn}
\usepackage{bm}
\usepackage{hyperref}
\usepackage{xcolor}
\usepackage[normalem]{ulem}

\newcommand{\Geffsq}[0]{G^2_\mathrm{eff}}
\newcommand{\Geff}[0]{G_\mathrm{eff}}

\newcommand{\mnras}{{MNRAS}}

\begin{document}

\preprint{APS/123-QED}

\title{The two-mode puzzle:\\Confronting self-interacting neutrinos with the full shape of the galaxy power spectrum} 

\author{David Camarena}
 \email{dcamarena93@unm.edu}
\author{Francis-Yan Cyr-Racine}%
\author{John Houghteling}
\affiliation{%
 Department of Physics and Astronomy,\\ University of New Mexico, Albuquerque, New Mexico 87106, USA
}%

\date{\today}

\begin{abstract}

A cosmological scenario in which the onset of neutrino free streaming in the early Universe is delayed until close to the epoch of matter-radiation equality has been shown to provide a good fit to some cosmic microwave background (CMB) data, while being somewhat disfavored by Planck CMB polarization data. To clarify this situation, we investigate in this paper CMB-independent constraints on this scenario from the Full Shape of the galaxy power spectrum. Although this scenario predicts significant changes to the linear matter power spectrum, we find that it can provide a good fit to the galaxy power spectrum data. Interestingly, we show that the data display a modest preference for a delayed onset of neutrino free streaming over the standard model of cosmology, which is driven by the galaxy power spectrum data on mildly non-linear scales. This conclusion is supported by both profile likelihood and Bayesian exploration analyses, showing robustness of the results. Compared to the standard cosmological paradigm, this scenario predicts a significant suppression of structure on subgalactic scales. While our analysis relies on the simplest cosmological representation of neutrino self-interactions, we argue that this persistent --- and somehow consistent --- picture in which neutrino free streaming is delayed motivates the exploration of particle models capable of reconciling all CMB, large-scale structure, and laboratory data.
\end{abstract}

\maketitle


\section{\label{sec:introd}Introduction}

Although neutrinos are the least understood particles in the Standard Model (SM), they play a crucial role in the evolution of the Universe. Since they couple gravitationally to everything else, their presence impacts cosmological observables on a broad range of scales, leading to observational features that can be used to constrain some of their hitherto unknown properties. For instance, one can use cosmological data to provide important --- and competitive --- constraints on the sum of the masses of neutrinos~\cite[see e.g.~Refs.~][]{Planck:2018vyg,Lattanzi:2017ubx,Sakr:2022ans}.

Besides their mass, cosmological observables can be also used to study new interactions in the neutrino sector. In the SM picture, neutrinos decouple and begin to free streaming when the temperature of the cosmic plasma drops to $\sim 1.5$~MeV. Since they are still gravitationally coupled, the free-streaming neutrinos effectively tug the baryon-photon fluid modifying the evolution of the cosmological perturbations. Such modifications, which appear as a phase shift and suppression of the amplitude of the cosmic microwave background (CMB) power spectra~\cite{Bashinsky:2003tk,Baumann:2015rya}, have been used to constrain the nature of the free streaming of neutrinos~\cite{Trotta:2004ty,Melchiorri:2006xs,DeBernardis:2008ys,Smith:2011es,Archidiacono:2011gq,Archidiacono:2012gv,Gerbino:2013ova,Archidiacono:2013fha,Melchiorri:2014yoa,Sellentin:2014gaa,Planck:2013pxb} through the $c_\mathrm{eff}$ and $c_\mathrm{vis}$ parametrization~\cite{Hu:1998kj}. 

Nonetheless, the presence of new physics in the neutrino sector can significantly alter the onset of the neutrino free streaming and, therefore, leave particular imprints in the Universe that can not simply be modeled by the $c_\mathrm{eff}$ and $c_\mathrm{vis}$ fluid approximation, calling for the necessity of considering a more realistic physical representation of the neutrino decoupling~\cite{Cyr-Racine:2013jua,Oldengott:2017fhy}. Thus, in the last few years, several models with non-standard neutrinos have been considered to study the cosmological consequences of altering the free streaming of neutrinos~\cite[see Refs.~][for instance]{Konoplich:1988mj,Berkov:1988sd,Belotsky:2001fb,Cyr-Racine:2013jua,Archidiacono:2013dua,Lancaster:2017ksf,Oldengott:2017fhy,Choi:2018gho,Song:2018zyl,Lorenz:2018fzb,Barenboim:2019tux,Forastieri:2019cuf,Smirnov:2019cae,Escudero:2019gvw,Ghosh:2019tab,Funcke:2019grs,Sakstein:2019fmf,Mazumdar:2019tbm,Blinov:2020hmc,deGouvea:2019qaz,Froustey:2020mcq,Babu:2019iml,Kreisch:2019yzn,Park:2019ibn,Deppisch:2020sqh,Kelly:2020pcy,EscuderoAbenza:2020cmq,He:2020zns,Ding:2020yen,Berbig:2020wve,Gogoi:2020qif,Barenboim:2020dmg,Das:2020xke,Mazumdar:2020ibx,Brinckmann:2020bcn,Kelly:2020aks,Esteban:2021ozz,Arias-Aragon:2020qip,Du:2021idh,CarrilloGonzalez:2020oac,Huang:2021dba,Sung:2021swd,Escudero:2021rfi,RoyChoudhury:2020dmd,Carpio:2021jhu,Orlofsky:2021mmy,Green:2021gdc,Esteban:2021tub,Venzor:2022hql,Taule:2022jrz,RoyChoudhury:2022rva,Loverde:2022wih,Kreisch:2022zxp,Das:2023npl,Venzor:2023aka,Sandner:2023ptm}.

Interestingly, the analyses of self-interacting neutrino models have unveiled that some CMB data allow for a significant delay in the onset of the free streaming of neutrinos~\cite{Cyr-Racine:2013jua,Archidiacono:2013dua,Lancaster:2017ksf,Oldengott:2017fhy,Barenboim:2019tux,Das:2020xke,Mazumdar:2019tbm,Brinckmann:2020bcn,RoyChoudhury:2020dmd,Kreisch:2019yzn,Park:2019ibn,Kreisch:2022zxp,Das:2023npl} and agree with two divergent pictures of the Universe: i) a paradigm where neutrinos moderately interact (MI$_\nu$) --- cosmologically reassembling the SM neutrinos, and ii) a cosmological picture where neutrinos strongly interact among themselves (SI$_\nu$). On the other hand, Planck CMB polarization data \cite{Planck:2018vyg} seems to disfavor the simplest representation of the SI$_\nu$ mode~\cite{Das:2020xke,Mazumdar:2020ibx,Brinckmann:2020bcn,RoyChoudhury:2020dmd}, which contrasts with data from the Atacama Cosmology Telescope (ACT) \cite{ACT:2020gnv}, which tends to favor the SI$_\nu$ mode~\cite{Kreisch:2022zxp,Das:2023npl}. Moreover, analyses of the phase of CMB peaks \cite{Follin:2015hya} and of the baryon acoustic oscillation (BAO) \cite{Baumann:2017lmt,Baumann:2019keh} have shown consistency with the expected phase shift from SM free-streaming neutrinos, further complicating the picture. 

The existence of the SI$_\nu$, which was first reported almost a decade ago~\cite{Cyr-Racine:2013jua,Archidiacono:2013dua} and has persisted in various comprehensive analyses, including those using the most recent cosmological data~\cite{Kreisch:2022zxp,Das:2023npl}, has so far been entirely driven by CMB data, with little information about the large-scale structure (LSS) of the Universe included in these analyses (beyond the BAO geometric distances and CMB lensing). 

This is significant as, due to a difference in the amplitude, $A_{\rm s}$, and tilt, $n_{\rm s}$, of the primordial curvature power spectrum favored by the SI$_\nu$, this alternate cosmological scenario predicts conspicuous changes to the linear matter power spectrum \cite{Kreisch:2019yzn} that could significantly impact the LSS of the Universe.

In this work, we investigate how a delayed onset of neutrino free streaming produced by the presence of novel self-interactions impacts the LSS of the Universe. Using the so-called Full Shape of the galaxy power spectrum~\cite{Ivanov:2019pdj,DAmico:2019fhj} along with Big Bang Nucleosynthesis (BBN) data \cite{Cooke:2017cwo,Aver:2015iza} and an effective four-fermion interaction to model self-interacting neutrinos, we show below, for the first time, that the large-scale distribution of galaxies displays a modest preference for new interactions in the neutrino sector. Crucially, this preference points to the same interaction strength favored by some CMB data, indicating that the evidence for new neutrino interactions is not likely driven by fortuitous noise features in these data sets. Since we focus here on the impact of neutrino interactions on LSS, we keep our analyses CMB agnostic. We will explore in an upcoming work whether self-interacting neutrinos create a statistically consistent scenario for both CMB and LSS data at the same time \cite{Camarena:2023bbb} (see also Ref.~\cite{He:2023oke} for a similar recent analysis).

This paper is organized as follows. In Sec.~\ref{sec:model},
we present the phenomenological neutrino interaction model used here and discuss its cosmological implications at linear scales. We then discuss the imprints that self-interacting neutrinos leave on the galaxy power spectrum in Sec.~\ref{sec:FS}. The data and methodology used in this paper are presented in Sec.~\ref{sec:data}, while our results and discussion are shown in Sec.~\ref{sec:result}. Finally, we conclude in Sec.~\ref{sec:conclu}. 

\section{\label{sec:model}Phenomenological model of self-interacting neutrinos}

Novel neutrino self-interactions beyond the SM delay the onset of their free streaming, hence suppressing the only source of anisotropic stress at early times. This suppression, in turn, impacts the evolution of the gravitational potentials, leaving significant modifications on both the evolution of photon and matter fluctuations~\cite{Bashinsky:2003tk,Baumann:2015rya}. From the cosmological point of view, a delayed onset of neutrino free streaming can be phenomenologically embodied by the simplest representation of self-interacting neutrinos: an effective four-fermion interaction characterized by a dimensionful Fermi-like constant~$\Geff$ coupling universally to all neutrino flavors. This leads to an interaction rate of the form 
\begin{eqnarray}\label{eq:Gamma_nu}
    \Gamma_\nu \equiv aG^2_{\mathrm{eff}}T^5_{\nu} \,,
\end{eqnarray}
with $T_\nu$ being the background temperature of neutrinos, and $a$ is the scale factor describing the expansion of the Universe. 

As shown in Ref.~\cite{Kreisch:2019yzn}, this simple representation can indeed serve as a proxy to study the changes that a delayed onset of neutrino free streaming produces on cosmological observables. However, we stress that Eq.~\eqref{eq:Gamma_nu}, which can be thought of as arising from neutrinos universally interacting via a massive mediator (see e.g.~Ref.~\cite{Berryman:2022hds} for a review), is unlikely to correspond to a realistic configuration of self-interacting neutrinos. Indeed, when taken  at face value, results from the study of supernovae~\cite{Kolb:1987qy,Manohar:1987ec,Dicus:1988jh,Davoudiasl:2005fd,Sher:2011mx,Fayet:2006sa,Choi:1989hi,Blennow:2008er,Galais:2011jh,Kachelriess:2000qc,Farzan:2002wx,Zhou:2011rc,Jeong:2018yts,Chang:2022aas} (see also Refs.~\cite{Fiorillo:2023cas,Fiorillo:2023ytr}); BBN~\cite{Ahlgren:2013wba,Huang:2017egl,Venzor:2020ova}; IceCube experiments~\cite{Ng:2014pca,Ioka:2014kca,Cherry:2016jol}; particles colliders~\cite{Bilenky:1992xn,Bardin:1970wq,Bilenky:1999dn,Brdar:2020nbj,Lyu:2020lps}; and decay kinematics of meson, leptons, tritium, and gauge boson~\cite{Brdar:2020nbj,Lyu:2020lps,Lessa:2007up,Bakhti:2017jhm,Arcadi:2018xdd,Blinov:2019gcj} exclude the flavor-universal parameter space of $\Geff$ capable of modifying the evolution of perturbations. Additionally, we note that in this simple flavor-independent framework, the SI$_\nu$ mode is significantly disfavored by the Planck polarization data~\cite{Das:2020xke,Mazumdar:2020ibx,Brinckmann:2020bcn,RoyChoudhury:2020dmd}. Taken together, these constraints indicate that more complex flavor-dependent interactions are very likely required to realize a viable model. Nonetheless, since current data on the large-scale distribution of galaxies do not yet have the same constraining power as the CMB and are thus unlikely to be sensitive to the minute details of the interactions, we perform our analysis here using the flavor-universal rate given in Eq.~\eqref{eq:Gamma_nu}. The use of this model also has the advantage of allowing for a direct comparison with previous results, including those obtained from the analysis of data from the ACT \cite{Kreisch:2022zxp}.

Besides using $\Geff$ to control the onset of neutrino free streaming, we also consider the effective number of relativistic species, $N_\mathrm{eff}$, as a free parameter of the model. For the sake of simplicity, we fix the total mass of neutrinos to $\Sigma m_{\nu} = 0.06$ eV and assume a single massive neutrino containing all the mass instead of several degenerate massive neutrinos. Although $\Sigma m_\nu$ has a crucial role in the analysis of CMB data, we note that current LSS data only weakly constrain this parameter \cite{Ivanov:2019hqk,Colas:2019ret}. Therefore, including neutrino masses in the analysis will only increase the uncertainties of our final results. Yet, as shown in appendix~\ref{app:mass}, the assumption of a fixed mass does not affect our conclusions. We stress that $\Geff$ is given in units of MeV$^{-2}$, and that the usual Fermi constant corresponds to the value $\Geff \sim \mathcal{O}(10^{-11})$ MeV$^{-2}$. Unless otherwise stated, we assume here that models laying at $-5.5 \leq \log_{10}(\Geff/\mathrm{MeV}^{-2}) \leq -2.5$ belong to the MI$_{\nu}$, and cosmologies following $ -2.5 < \log_{10}(\Geff/\mathrm{MeV}^{-2}) \leq 0.5$ are associated to the SI$_{\nu}$ regime. 

To offer a fair comparison (in terms of degrees of freedom) between our case of study and the typical picture of neutrinos starting to free stream around $T \sim 1.5$ MeV, here, we use the $\Lambda\mathrm{CDM} + N_\mathrm{eff}$ model (with two massless and one massive neutrino with $\Sigma m_{\nu} = 0.06$ eV) to represent the standard paradigm. When necessary, we use the subscript (overscript) $I_\nu$ to denote self-interacting neutrinos quantities, while simply using $\Lambda$CDM to denote quantities related to the $\Lambda\mathrm{CDM} + N_\mathrm{eff}$ model.

\subsection{\label{sub:boltzmann}Collision term and Boltzmann equations}
As is usually done for ultra-relativistic particles \cite{Ma:1995ey}, we expand the scalar neutrino temperature fluctuations with wavenumber ${\bf k}$, proper momentum ${\bf p}$, and conformal time $\tau$ in terms of Legendre polynomial $P_\ell$ as
\begin{equation}
    \frac{\delta T_\nu}{T_\nu}({\bf k},{\bf p},\tau) = \frac{1}{4}\sum_{\ell = 0}^\infty (-i)^\ell (2\ell+1) \nu_\ell(k,p,\tau)P_\ell(\mu),
\end{equation}
where $p = |\mathbf{p}|$, $k = |{\bf k}|$, and $\mu$ is the cosine of the angle between ${\bf k}$ and ${\bf p}$. Using the self-interaction rate given in Eq.~\eqref{eq:Gamma_nu} and the above decomposition, we can compute the collision term entering in the right-hand side of the Boltzmann equations to later derive the set of equations that will describe the evolution of cosmological perturbations in the presence of self-interacting neutrinos. Under the thermal approximation, the collision term at first order for the $\nu \nu \rightarrow \nu \nu$ process is given by \cite{Kreisch:2019yzn}
\begin{eqnarray} \label{eq:collision}
    C_{\nu} \left[ \mathbf{p} \right] && = \frac{\Geffsq T_\nu^6}{4} \frac{\partial \ln f_\nu^{(0)}}{\partial \ln p} \sum_{\ell = 0}^{\infty} (-i)^{\ell} (2\ell + 1) \nu_\ell P_\ell(\mu) \nonumber \\
    && \times \left[ A\left(\frac{p}{T_\nu}\right) + B_\ell\left(\frac{p}{T_\nu}\right) - 2D_\ell\left(\frac{p}{T_\nu}\right) \right] \,,
\end{eqnarray}
where $f^{(0)}_\nu$ is the background (Fermi-Dirac) neutrino distribution function, and $A(x)$, $B_\ell(x)$, and $C_\ell(x)$ are functions related to the different integral terms in the collision term \cite[see App.~C~and~D in Ref.][for a detailed derivation of the collision term]{Kreisch:2019yzn}.

Adopting the conformal Newtonian gauge, and using the collision term defined above, we derive the Boltzmann equations for massive neutrinos
\begin{eqnarray}\label{eq:boltzmann_massive}
    \frac{\partial \nu_\ell}{\partial \tau}  && = - \frac{kq}{\epsilon} \left( \frac{\ell + 1}{2\ell + 1} \nu_{\ell + 1} - \frac{\ell}{2\ell + 1} \nu_{\ell -1} \right) + 4 \left[\frac{\partial \phi}{\partial \tau} \delta_{\ell 0} \right. \nonumber \\
    &&  \left. + \frac{k}{3}\frac{\epsilon}{q} \psi \delta_{\ell 1}  \right] - \frac{\Gamma_\nu}{f_\nu^{(0)}} \left(\frac{T_{\nu,0}}{q}\right) \left[A\left(\frac{q}{T_{\nu,0}}\right) \right. \nonumber \\  
    && \left. + B_\ell\left(\frac{q}{T_{\nu,0}}\right) -2D_\ell\left(\frac{q}{T_{\nu,0}}\right) \right]\nu_{\ell} \,,
\end{eqnarray}
 where $\phi$ and $\psi$ are the scalar perturbations of the conformal Newtonian gauge, $\nu_\ell$ is typical perturbation variable expanded in Legendre polynomials, $T_{\nu,0}$ is the current temperature of neutrinos, $q = a p$ is the comoving momentum, and $\epsilon = \sqrt{q^2 + a^2 m^2_{\nu}}$ with $m_\nu$ being the mass of the neutrino species. 

Analogously, the Boltzmann equations for massless neutrinos can be derived by setting $\epsilon=q$ and averaging Eq.~\eqref{eq:boltzmann_massive} over momentum with $f^{(0)}_\nu$, 
\begin{eqnarray}\label{eq:boltzmann_massless}
    \frac{\partial F_\ell}{\partial \tau} && = -k \left( \frac{\ell + 1}{2\ell + 1} F_{\ell + 1} - \frac{\ell}{2\ell + 1} F_{\ell -1} \right) \\
    && + 4 \left[\frac{\partial \phi}{\partial \tau} \delta_{\ell 0} + \frac{k}{3}\psi \delta_{\ell 1} \right] - \alpha_\ell  \Gamma_\nu F_\ell \,,
\end{eqnarray}
where $F_\ell$ is the perturbation variable for massless neutrinos as defined in Ref.~\cite{Ma:1995ey} and the collision term is characterized by:
\begin{equation}\label{eq:alpha}
    \alpha_\ell = \frac{120}{7\pi^4} \int^{\infty}_{0} \mathrm{d}x\: x^2 \left[A(x) + B_\ell(x) - 2 D_\ell(x)\right] \,.
\end{equation}
We highlight that Eqs.~\eqref{eq:boltzmann_massive}~and ~\eqref{eq:boltzmann_massless} satisfy the energy and momentum conservation since the collision terms follow the relation $\alpha_{\ell} = A + B_{\ell} - 2 D_{\ell} = 0$ for $\ell = \left\lbrace 0,1\right\rbrace$.

We implemented Eqs.~\eqref{eq:boltzmann_massive} and \eqref{eq:boltzmann_massless} in the cosmological code \texttt{CLASS-PT}~\cite{Blas:2011rf,Chudaykin:2020aoj}, which uses the Eulerian perturbations theory to compute the galaxy power spectrum at mildly non-linear scales. To avoid the stiffness of the Boltzmann equations at early times, when the mean free path of self-interacting neutrinos is much smaller than the Hubble horizon, we use the so-called tight-coupling approximation~\cite{Cyr-Racine:2010qdb}. More precisely, we use the tight-coupling approximation at times where $\Gamma_{\nu} > 10^{3} \mathcal{H}$, with $\mathcal{H}$ being the Hubble rate at conformal time~\footnote{Our modified version of \texttt{CLASS-PT} as well as a more detailed description of our numerical implementation are available at \url{https://github.com/davidcato/class-interacting-neutrinos-PT}.}.

\subsection{\label{sub:cosmo} Cosmological implications of a delayed free streaming}

In this Section, we briefly review how a delay in the free streaming of neutrinos impacts the cosmological observables, namely, the CMB and linear matter power spectrum. We refer the reader to Ref.~\cite{Kreisch:2019yzn} for a thorough discussion of how the assumption of Eq.~\eqref{eq:Gamma_nu} impacts the evolution of cosmological perturbations.
All models discussed in this section use the same values for the cosmological parameters, except for $\Geff$, as well as $A_{\rm s}$ and $n_{\rm s}$ when indicated, whose values vary according to what is indicated in the corresponding label. 

The changes that self-interacting neutrinos produce in the CMB power spectrum can be explained in terms of the gravitational pull produced by free-streaming radiation species. In the standard paradigm, after their decoupling, neutrinos travel supersonically across the Universe, gravitationally pulling the photon-baryon wave toward larger scales~\cite{Bashinsky:2003tk}. 
This gravitational tug felt by the photon-baryon wave results in a phase shift toward smaller~$\ell$ --- larger scales --- and a reduction of the amplitude of the CMB power spectra~\cite{Bashinsky:2003tk,Cyr-Racine:2013jua,Baumann:2015rya}. Contrastingly, a delay in the free streaming of neutrinos boosts the amplitude of the CMB power spectra and leads to a phase shift toward larger~$\ell$ --- smaller scales. This behavior also manifests as a small reduction of the sound horizon scale of photons, which marginally help to accommodate larger values  of $H_0$ in the CMB power spectrum~\cite[see Ref.][for instance]{Brinckmann:2020bcn}.

On the other hand, the changes that self-interacting neutrinos imprint on the evolution of dark matter fluctuations, and consequently, on the matter power spectrum, are better understood by examining the gravitational potentials $\psi$ and $\phi$ in Newtonian gauge. Delaying the onset of the neutrino free streaming suppresses the anisotropic stress of the Universe, altering the evolution of the gravitational potentials by setting $\psi = \phi$ until the onset of neutrino free streaming. Effectively, this suppression increases the initial value of the gravitational potential $\psi$ and enhances its oscillatory envelope at horizon entry~\cite{Kreisch:2019yzn}. 
Depending on the scale, the interplay of these effects results in either a faster or slower decay of the gravitational potential $\psi$ in comparison to the $\Lambda$CDM model. This feature gives rise to scale-dependent behavior of the dark matter perturbations, which can be distinguished by observing three different kinds of Fourier modes: $k^{\rm tc}_{\rm h}$, a mode entering the horizon while neutrinos are still tightly coupled; $k^{\rm fs}_{\rm h}$, a mode entering the horizon when neutrinos start to free stream; and $k_{\rm h}$, a mode that crosses the horizon well after the onset of the neutrino free streaming.

Dark matter perturbations entering the horizon while neutrinos are still tightly coupled, $k^{\rm tc}_{\rm h}$, will undergo an initial enhancement in amplitude at horizon entry due to an increase in the initial value of the gravitational potential $\psi$. However, the absence of anisotropic stress will also amplify the oscillatory envelope of $\psi$, leading to slower decay of the gravitational potential and, consequently, resulting in a net damping of amplitude of the dark matter perturbations compared to the $\Lambda$CDM picture. On the other hand, modes entering the horizon when the free streaming begins, $k^{\rm fs}_{\rm h}$, will be influenced by the change in the initial conditions and the faster decay of $\psi$, implying an enhancement in the dark matter perturbation amplitude~\cite{Kreisch:2019yzn}. Conversely, modes entering the horizon well after the beginning of neutrino free streaming, $k_{\rm h}$, will remain unaltered compared with the standard cosmological picture.

\begin{figure}
\includegraphics[width=0.995\columnwidth]{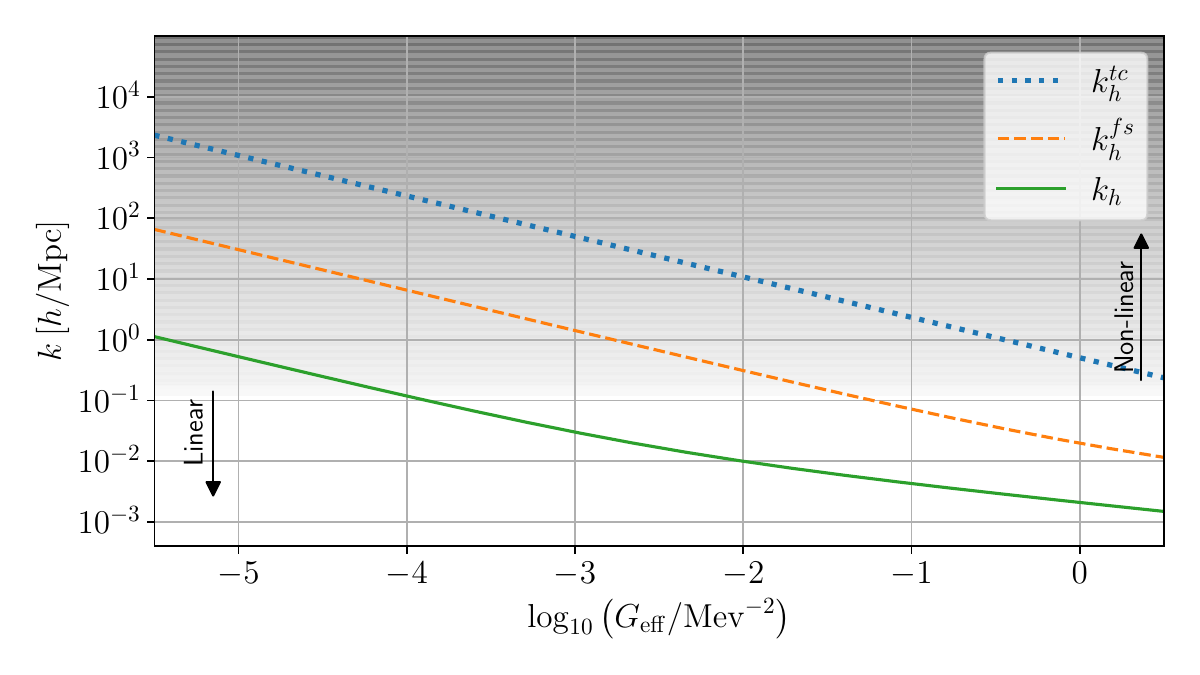}
\caption{\label{fig:k_Geff} Modes crossing the horizon during the neutrino tight-coupling era, $k^{\rm tc}_{\rm h}$, the onset of free streaming, $k^{\rm fs}_{\rm h}$, and well after the self-decoupling, $k_{\rm h}$, as functions of the self-interaction coupling $\Geff$. The gray gradient represents the transition between the linear and non-linear scales at redshift 0. As described in the main text, self-interacting neutrinos lead to substantial changes in the evolution of dark matter perturbations at modes $k_{\rm h}^{\rm fs}$ and $k_{\rm h}^{\rm tc}$. For a strongly interacting neutrino model with $\log_{10} (\Geff / \mathrm{MeV}^{-2}) \sim -1.5$, these modes will belong to the (mildly) non-linear realm today.}
\end{figure}

 Figure~\ref{fig:k_Geff} shows the typical order of magnitude of the aforementioned modes as functions of the coupling strength~$\Geff$. An extreme delay to the onset of free streaming produced by strongly self-interacting neutrinos, for instance, $\log_{10} (\Geff/\mathrm{MeV}^{-2}) \sim 0$, will lead to substantial changes in the evolution of perturbations on linear scales $k_{\rm h}^{\rm fs} \sim 0.05 \: h/\mathrm{Mpc}$ (in this paper, we always refer to scales being linear or non-linear at $z = 0$). Meanwhile, a less extreme SI$_{\nu}$ model with $\log_{10} (\Geff/\mathrm{MeV}^{-2}) \sim -2$ will mainly modify the evolution of perturbations in the (mildly) non-linear regime $k_{\rm h}^{\rm fs} \sim 0.5 \: h/\mathrm{Mpc}$. Furthermore, cosmologies well inside the MI$_\nu$ regime, for instance $\log_{10} (\Geff/\mathrm{MeV}^{-2}) \sim -4$, will induce changes at highly non-linear scales. 

 \begin{figure}
\includegraphics[width=0.995\columnwidth]{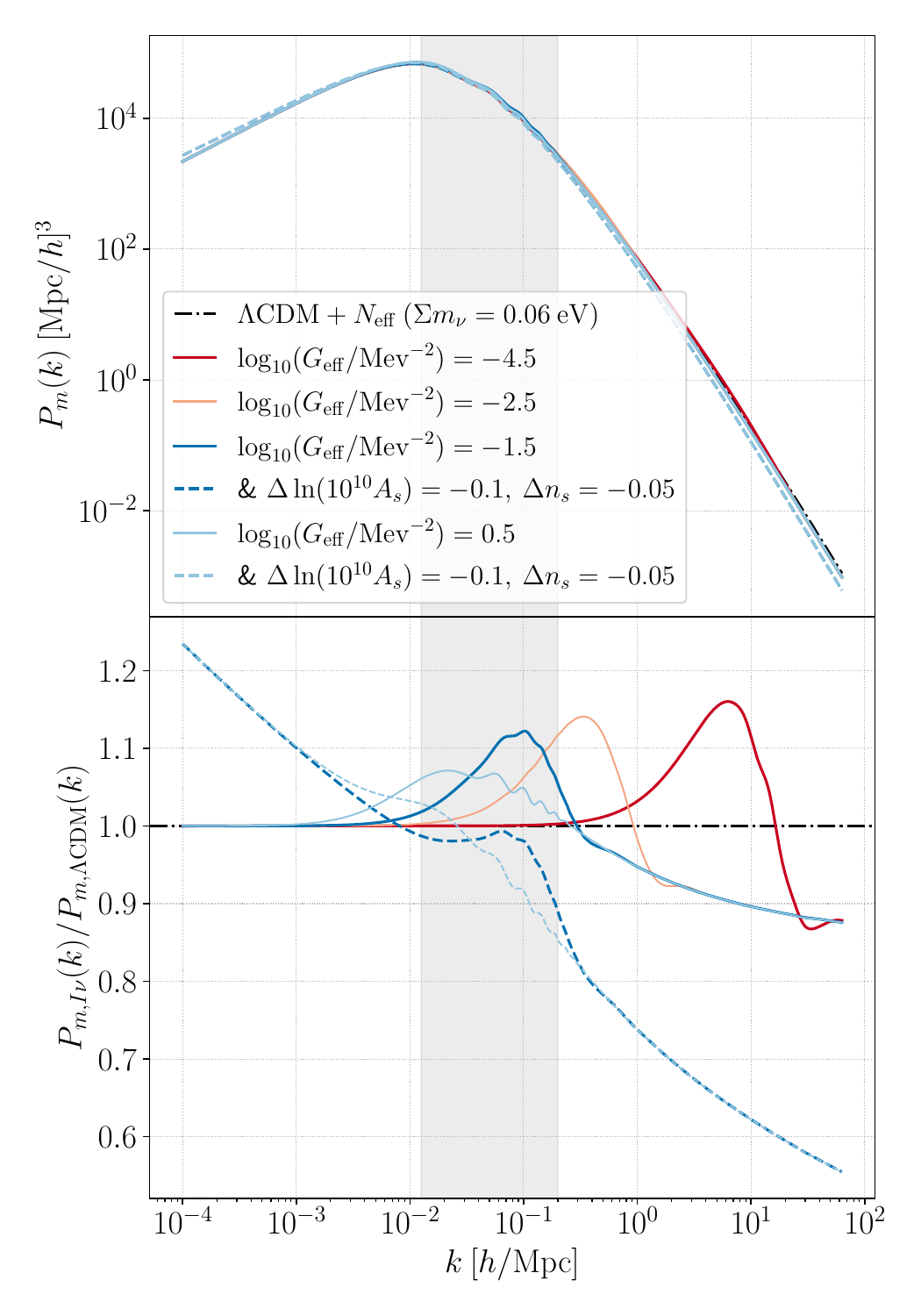}
\caption{\label{fig:Pk_lin_Geff_vs_BOSS-BF} Linear matter power spectrum (top panel) for the $\Lambda\mathrm{CDM} + N_\mathrm{eff}$ cosmology and the self-interacting neutrino model with different values of $\Geff$. The ratios of the latter with the $\Lambda\mathrm{CDM} + N_\mathrm{eff}$ model are shown in the bottom panel. Dashed and solid lines of the same color use the same value of $\Geff$ but lower values of $A_{\rm s}$ and $n_{\rm s}$ as specified in the labels. A delay in the free streaming of neutrinos driven by $\log_{10} (\Geff/\mathrm{MeV}^{-2}) = -1.5$ (solid dark blue line) suppresses and enhances the power spectrum at $k^{\rm tc}_{\rm h} \sim 10 \: h/\mathrm{Mpc}$ and $k^{\rm fs}_{\rm h} \sim 0.2 \: h/\mathrm{Mpc}$, respectively. A decrease in $A_{\rm s}$ and $n_{\rm s}$ (dashed blue lines), however, smooths out the bump observed around $k^{\rm fs}_{\rm h}$ and leads to a red-tilted power spectrum. The dashed dark blue line approximately corresponds to the SI$_\nu$ mode. The gray band represents the range of scales probed by the galaxy power spectrum data.}
\end{figure}

\begin{figure*}[t]
\includegraphics[width=0.95\textwidth]{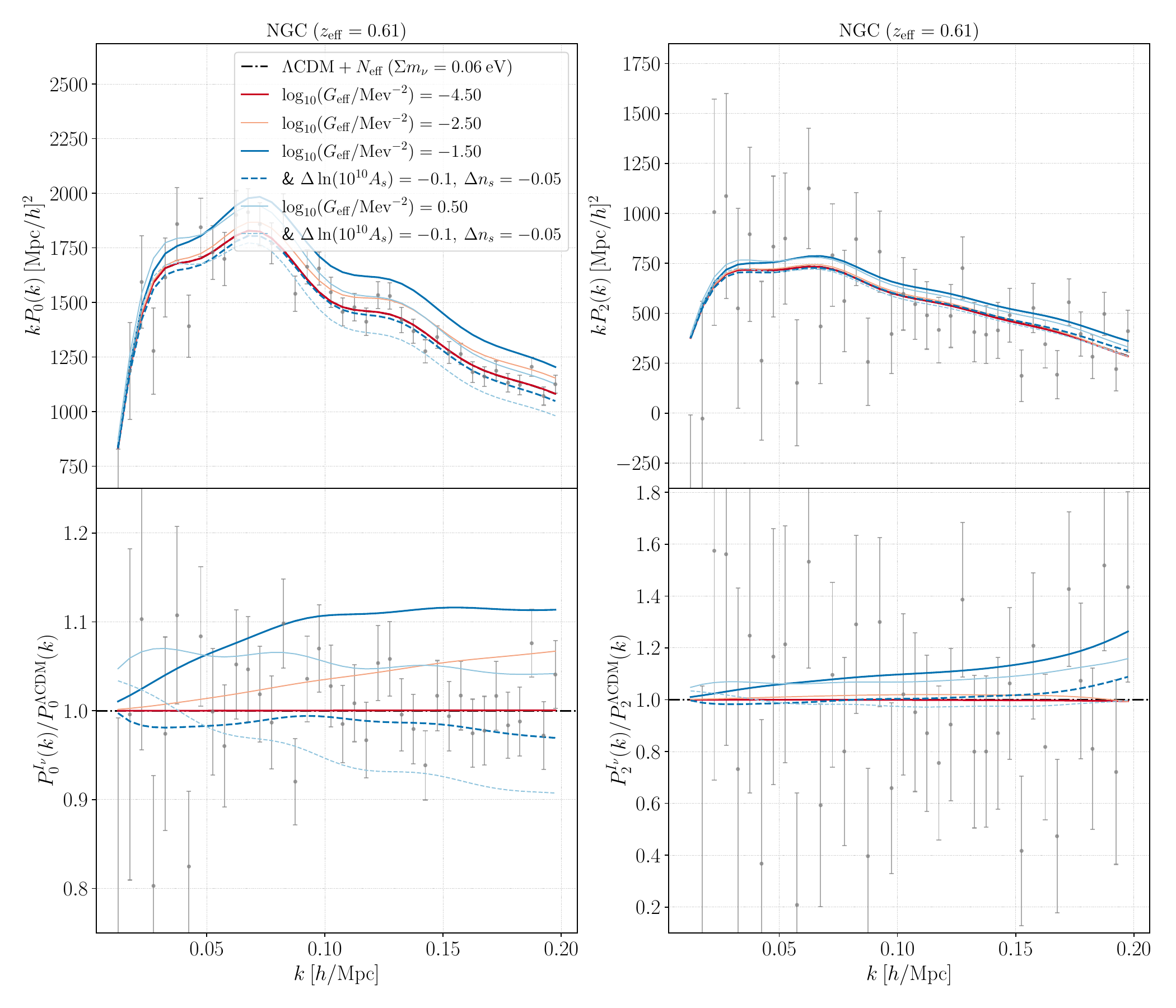}
\caption{\label{fig:multipoles_Geff_vs_BOSS-BF_high-z_NGC} Impact of $\Geff$ in the monopole (left) and quadropole (right) of the galaxy power spectrum. The dashed and solid lines of the same color represent models with the same value of $\Geff$ but lower values of $A_{\rm s}$ and $n_{\rm s}$. Models well inside the MI$_\nu$ regime results indistinguishable from the $\Lambda\mathrm{CDM} + N_\mathrm{eff}$ case, however, sizable deviations are produced by models in the SI$_\nu$ regime. Interestingly, a decrease in $A_{\rm s}$ and $n_{\rm s}$ can potentially compensate for the impact of $\Geff$ (dashed dark blue line). This illustrates that the so-called SI$_\nu$ mode could offer a good fit to the current LSS data. The data displayed here correspond to the subset in the NGC at $z_\mathrm{eff} = 0.61$ (see Sec.~\ref{sec:data}).}
\end{figure*}

Figure~\ref{fig:Pk_lin_Geff_vs_BOSS-BF} shows the matter linear power spectrum for the $\Lambda$CDM + $N_\mathrm{eff}$ model and the self-interacting neutrino scenario with different values of $\Geff$; the gray band illustrates the Fourier modes probed through the full shape of the galaxy power spectrum data. We note that a delay of the free streaming induced by $\log_{10} (\Geff/\mathrm{Mev}^{-2}) \approx -2.5$ (light salmon line) suppresses the linear power spectrum up to a factor of $\gtrsim 10\%$ at scales $k^{\rm tc}_{\rm h} \sim 20\: h/\mathrm{Mpc}$ while enhancing by overall the same factor modes around $k_{\rm h}^{\rm fs} \sim 0.5\: h/\mathrm{Mpc}$. The power spectrum for modes crossing the horizon well after the onset of free streaming, here $k_{\rm h} \lesssim 0.01\: h/\mathrm{Mpc}$, remains unaltered compared to the $\Lambda$CDM + $N_\mathrm{eff}$ model. Due to a reduction of the radiation energy density, scenarios in which the free streaming of neutrinos starts after the matter-radiation equality experience a lower enhancement in the power spectrum at scales $k_{\rm h}^{\rm fs}$. This is illustrated by $\log_{10}(\Geff/\mathrm{Mev}^{-2}) \approx 0.5$ (solid light blue line), where one can note that the power spectrum just increases by a factor of $\sim 8.5\%$ on scales $k_{\rm h}^{\rm fs} \approx 0.03 \:h/\mathrm{Mpc}$. Analogously to the CMB, self-interacting neutrinos induce a phase shift in the matter power spectrum around the typical BAO scale; this is $k \approx 0.1 \:h/\mathrm{Mpc}$. This shift is particularly appreciable for models that delay the onset of free-streaming neutrinos until close to recombination; see, for instance, the case of $\log_{10}(\Geff/\mathrm{Mev}^{-2}) \approx 0.5$ (solid and dashed light blue lines). 

We recall that the SI$_\nu$ mode offers a good fit to the CMB data at the cost of reducing the amplitude, $A_{\rm s}$, and tilt, $n_{\rm s}$, of the primordial scalar power spectrum~\cite{Kreisch:2019yzn,Taule:2022jrz,Das:2020xke,Oldengott:2017fhy}. The dashed blue lines in Fig.~\ref{fig:Pk_lin_Geff_vs_BOSS-BF} show that a decrease in $A_{\rm s}$ and $n_{\rm s}$ not only produces a red-tilted power spectrum but also eases the bump expected on scales $k^{\rm fs}_{\rm h}$. We note that the~SI$_\nu$-like~mode (dashed dark blue line) produces substantial changes across the linear and non-linear scales --- several of those scales will be accessible through the multipoles of the galaxy power spectrum (gray band).

\section{\label{sec:FS} Full shape of the galaxy power spectrum}

Depending on the value of the coupling strength, self-interacting neutrinos leave particular imprints either at linear and/or (mildly) non-linear scales today. As shown by Fig.~\ref{fig:k_Geff} the most relevant cosmological cases mainly generate changes at scales in which the linear perturbation theory breaks down, i.e., modes with $k \gtrsim 0.1\: h/\mathrm{Mpc}$. However, given that there is not a straightforward mapping between the linear and (mildly) non-linear power spectrum, it is complex to gauge \textit{a priori} the impact that the delaying of the onset of neutrinos will have in the (multipoles) galaxy power spectrum. 

In this Section, we explore how the deferring of the free streaming of neutrinos impacts the multipoles of the galaxy power spectrum. As stated before, we used a modified version of the publicly available \texttt{CLASS-PT} code~\cite{Blas:2011rf,Chudaykin:2020aoj}, which relies on the Eulerian perturbation theory and makes use of the Einstein-de Sitter (EdS) convolution kernels~\cite{Bernardeau:2001qr} to compute the galaxy power spectrum and its multipoles at one-loop correction. Given that one-loop redshift-space perturbation theory is expected to break down for modes beyond $k_\mathrm{max} \approx 0.25 \:h/\mathrm{Mpc}$~\cite{DAmico:2019fhj,Ivanov:2019pdj,Nishimichi:2020tvu}, we conservatively adopt $k_\mathrm{max} = 0.2\:h/\mathrm{Mpc}$ for our main analysis, although, complementary analysis exploring the impact of $k_\mathrm{max}$ will also be presented. Before examining the imprints that interacting neutrinos leave in the galaxy power spectrum, we qualitatively argue that the EdS kernels can be used even in the presence of self-interacting neutrinos as long as the onset of free streaming occurs before the matter-dominated era. 

Due to their supersonic velocity, massive neutrinos that become non-relativistic in the matter-dominated era do not cluster on scales lower than the so-called free-streaming scale, which is characterized by the wavenumber $k_\mathrm{NR} \approx 0.018 \:\Omega^{1/2}_{m,0} \left( m_{\nu}/1\:\mathrm{eV} \right)^{1/2} h/\mathrm{Mpc}$~\cite{Lesgourgues:2014zoa}. This produces a scale-dependent growth rate that leads to a  suppression of the linear power spectrum at modes $k > k_\mathrm{NR}$. This picture is expected to remain unchanged if the onset of the free streaming occurs before the matter-dominated era. Indeed, in such scenarios, the changes produced in the gravitational field $\psi$ at horizon crossing only modify the initial shape of the transfer function and do not introduce any additional features that alter the evolution of the perturbations in the non-linear regime. Since most of the $\Geff$ parameter space considered here delays the free streaming of neutrinos until close to the matter-dominated era, we argue that the mildly non-linear power spectrum can be computed using the EdS kernels. We also note that this is consistent with the fact that the self-interactions considered here do not modify the free-streaming scale to a very good approximation. 

The top panels of Fig.~\ref{fig:multipoles_Geff_vs_BOSS-BF_high-z_NGC} show the monopole (left) and quadrupole (right) of the galaxy power spectrum for the $\Lambda\mathrm{CDM} + N_\mathrm{eff}$ model and the self-interacting neutrino model with different values of $\Geff$. The ratios between the corresponding multipoles are shown in the bottom panels. All models shown here assume the same values for the galaxy power spectrum nuisance parameters. Concerning moderately self-interacting neutrinos, we note that models with a universal coupling of $\log_{10} (\Geff/\mathrm{MeV}^{-2}) = -4.5$ (red line) gives results indistinguishable from the standard picture, this being true both for the monopole and quadrupole. Meanwhile, models following $\log_{10} (\Geff/\mathrm{MeV}^{-2}) = -2.5$ produce a sizable enlargement of the monopole, without significantly deviating from the quadrupole predicted by the $\Lambda\mathrm{CDM} + N_\mathrm{eff}$ cosmology. Thus, similar to the case of CMB analysis, we expect the MI$_\nu$ regime and the standard cosmological model to provide a similar fit to the data. 

Additionally, Fig.~\ref{fig:multipoles_Geff_vs_BOSS-BF_high-z_NGC} shows that strongly interacting neutrinos (solid blue lines) significantly modify the different multipoles of the galaxy power spectrum. For instance, models with $\log_{10}(\Geff/\mathrm{MeV}^{-2}) = -1.5$ increase the amplitude of $P_0$ by a factor of $10\%$ on scales $k \gtrsim 0.08\: h/\mathrm{Mpc}$ while also departing from the quadrupole of the standard case as $k$ increases. Nonetheless, we note that a decrease in $A_{\rm s}$ and $n_{\rm s}$ reduces the offset between the SI$_\nu$ regime and the  $\Lambda\mathrm{CDM} + N_\mathrm{eff}$ model. This illustrates that the typical SI$_\nu$ mode found in CMB fits (dashed dark blue line) can potentially also offer a good fit to the galaxy power spectrum data.

\section{\label{sec:data} Data and methodology}

\subsection{\label{sub:fs} Full shape power spectrum and BAO}

We use the dataset from the twelfth data release of the Baryon Oscillation Spectroscopic Survey \cite{SDSS:2011jap,BOSS:2012dmf,BOSS:2016wmc} and its corresponding window-free galaxy power spectrum \cite{Philcox:2020vbm,Philcox:2021kcw} to constrain the presence of new interactions in the neutrino sector. The galaxies from BOSS DR12 are distributed across four different subsets, which correspond to two redshift slices, $0.2 < z < 0.5$ from the LOWZ sample ($z_\mathrm{eff} = 0.38$) and $0.5 < z < 0.75$ from the CMASS sample ($z_\mathrm{eff} = 0.61$), and two sky cuts in the north and south Galactic cap (NGC and SGC, respectively). The galaxy power spectrum data are given for each of these subsets. 

We constrain a potential delay in the onset of neutrino free streaming by analyzing the multipoles of the galaxy power spectrum $P_\ell(k,z)$ ($\ell = 0, 2, 4$)~\cite{Philcox:2021kcw,Chudaykin:2020ghx}, along with the $Q_0(k,z)$ estimator~\cite{Ivanov:2021fbu}.  This estimator, closely related to the real space power spectrum, is obtained using a linear combination of the first few power spectrum multipoles. As discussed in the previous Section, redshift-space perturbation theory breaks down for wavenumbers larger than $k_\mathrm{max} \approx 0.25\: h/\mathrm{Mpc}$. Thus, our main analysis conservatively uses the multipoles in the wavenumber range $k_\mathrm{min} = 0.01\: h$/Mpc and $k_\mathrm{max} = 0.2\: h$/Mpc. Since real-space perturbation theory can be safely applied to smaller scales \cite{Ivanov:2021fbu}, we consider measurements of the $Q_0$ metric in the range $k_\mathrm{min} = 0.2\: h$/Mpc and $k_\mathrm{max} = 0.4\: h$/Mpc. In both cases, we use a width bin of $\Delta k = 0.005\: h$/Mpc. Furthermore, we also use the reconstructed power spectrum, which provides constraints on the so-called Alcock-Paczynski (AP) parameters \cite{Philcox:2020vvt}.

We analyze this data using the BOSS likelihood presented in Ref.~\cite{Philcox:2021kcw}, which analytically marginalizes over the nuisance parameters that enter linearly in the power spectrum, i.e., the counterterms (monopole $c_0$, quadrupole $c_2$, hexadecapole $c_4$, and fingers-of-God $\tilde{c}$), the third-order galaxy bias $b_{\Gamma_3}$, and the stochastic contributions ($P_\mathrm{shot}$, $a_0$, and $a_1$). The covariance matrix used for this likelihood has been computed using MultiDark-Patchy 2048 simulations~\cite{Kitaura:2015uqa,Rodriguez-Torres:2015vqa}.

\subsection{\label{sub:bbn} Big Bang Nucleosynthesis}

Complementary to LSS data, we use BBN data to effectively constrain the baryon density parameter, $\omega_b$, and the effective number of relativistic species, $N_\mathrm{eff}$. In particular, we follow the implementation presented in Ref.~\cite{Schoneberg:2019wmt}, which uses an interpolation table that depends on $\omega_{\rm b}$ and $N_\mathrm{eff}$ (extracted from the PArthENoPe code \cite{Consiglio:2017pot}), along with a measurement of the nuclear rate $d(p,\gamma)^3\mathrm{He}$ \cite{Adelberger:2010qa}, to theoretically predict the primordial abundance of helium, $Y_\mathrm{He}$, and deuterium, $y_\mathrm{DP}$. We constrain the theoretically predicted values of the primordial abundance of helium and deuterium using the measurements presented by \citet{Cooke:2017cwo} and \citet{Aver:2015iza}, respectively. 

One might worry that neutrino self-interactions could impact BBN in a way that makes using the above prior inconsistent. However, neutrino self-interaction does not alter, in general, the standard electroweak decoupling of neutrinos from the rest of the Standard Model plasma. This is supported by the analysis presented in Ref.~\cite{Grohs:2020xxd}, which shows that the SI$_\nu$ scenario has only a very slight impact on the predicted primordial abundances of helium and deuterium. Thus, theoretical predictions from standard BBN can be safely used to constrain the self-interacting neutrino cosmology.

\subsection{\label{sub:stat} Scanning the Parameter Space}

Since the delay in the onset of neutrino free streaming is described by a single parameter, our model is specified by seven cosmological parameters: the self-interaction strength,  $\Geff$, and the six usual $\Lambda$CDM+$N_\mathrm{eff}$ parameters~\footnote{Since the LSS data we use are insensitive to reionization, we assume a fixed value for the optical depth $\tau_{\rm reio} = 0.05$.}. These latter parameters encompass the baryon density $\omega_{\rm b}$, the cold dark matter density $\omega_{\rm cdm}$, the Hubble constant $H_0$, the effective number of relativistic species $N_\mathrm{eff}$, and the amplitude and tilt of the primordial power spectrum, $A_{\rm s}$ and $n_{\rm s}$, respectively. In addition to this, the likelihood of the galaxy power spectrum includes three non-marginalized nuisance parameters for each subsample of the BOSS DR12, hence, twelve parameters are added to the parameter space. These parameters are the linear, $b_1$, quadratic, $b_2$, and second order galaxy, $b_{\Gamma_2}$, biases. 

We employ two different schemes to explore this nineteen-dimensional parameter space: a profile likelihood and a Metropolis-Hasting sampling. The frequentist approach, provided by the profile likelihood, will allow us to identify possible volume effects, better understand the goodness of fit of the model and obtain some useful examples to illustrate the physical predictions of different regions of the parameter space. The Bayesian exploration through a Metropolis-Hasting algorithm, on the other hand, will provide a broader understanding of the whole parameter space.

\begin{figure}[htb]
\includegraphics[width=0.95\columnwidth]{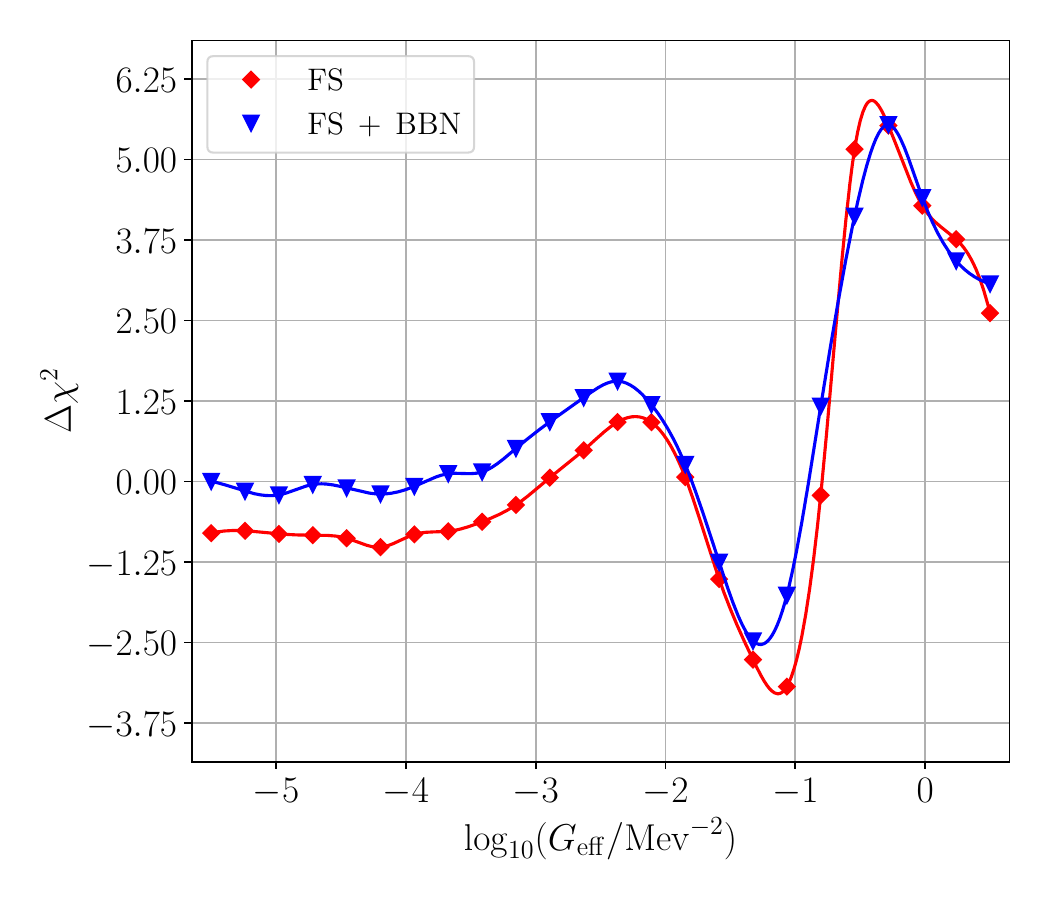}
\caption{\label{fig:profile_likelihood} Profile likelihood of the $\Geff$ parameter, expressed in terms of the goodness-of-fit of the self-interacting neutrino model $\Delta \chi^2 \equiv \chi^2_{\mathrm{min},I_{\nu}} - \chi^2_{\mathrm{min},\Lambda\mathrm{CDM}}$, for the analyses of FS and FS + BBN data. We note that models with $\log_{10} \left(\Geff/\mathrm{Mev}^{-2}\right) \approx -1.3$ provide a better fit to the data than the $\Lambda\mathrm{CMB}+N_\mathrm{eff}$ model.}
\end{figure}

\begin{table}[htb]
\caption{\label{tab:fiducial}
Best-fits to the FS+BBN data of the $\Lambda\mathrm{CDM}+N_\mathrm{eff}$ model and two cases of the self-interacting neutrino model that represent the MI$_\nu$ and SI$_\nu$ modes. While the MI$_\nu$ mode and the standard model offers a similar fit to the data, the SI$_\nu$ mode leads to a lower $\chi^2$ by decreasing $A_{\rm s}$ and $n_{\rm s}$.}
\begin{ruledtabular}
\begingroup
\renewcommand{\arraystretch}{1.35}
\begin{tabular}{lccc}
\hline
Parameter            &  $\Lambda \mathrm{CDM} + N_\mathrm{eff}$ &   MI$_\nu$ mode & SI$_\nu$ mode\\
\hline
$\log_{10}(\Geff/\mathrm{MeV}^{-2}) $ & - & $-4.98$ & $-1.33$ \\
$N_\mathrm{eff}$ & $2.94$ & $2.93$ & $2.95$ \\
$H_0$ [km s$^{-1}$ Mpc$^{-1}$] & $68.51$ & $68.21$ & $68.19$ \\
$\omega_{\rm b}$ & $0.0226$ & $0.0225$ & $0.0225$ \\
$\omega_{\rm cdm}$ & $0.129$ & $0.128$ & $0.126$ \\
$\ln(10^{10}A_{\rm s})$ & $2.85$ & $2.81$ & $2.74$ \\
$n_{\rm s}$ & $0.902$ & $0.906$ & $0.85$ \\
$\sigma_8$ & $0.762$ & $0.743$ & $0.720$ \\
$\chi^2_\mathrm{min}$ & $767.65$ & $767.44$ & $765.18$ \\
\hline
\end{tabular}
\endgroup
\end{ruledtabular}
\end{table}

We profile the likelihood in twenty-four different values of the self-interaction coupling constant. Such values are equally linearly spaced in the range $\log_{10}(\Geff/\mathrm{MeV}^{-2}) = [-5.5,0.5]$. We find the best-fit to the data for each particular value of $\Geff$ by minimizing the likelihoods with the Derivative-Free Optimizer for Least-Squares package~\cite{10.1145/3338517}. The main results of this analysis are a set of points that discretely sketch the profile likelihood. Additionally, we present a continuous representation of the profile likelihood. This representation is obtained applying a cubic spline to the $\chi^2_\mathrm{min}$ points found with the minimization algorithm. 

Complementary to this, and in order to unveil possible correlations between $\Geff$ and other cosmological parameters, we sample the parameter space using a Metropolis-Hasting algorithm. More specifically, we use the \texttt{montepython} code~\cite{Audren:2012wb,Brinckmann:2018cvx}. To avoid possible deficiencies in the sampling, we start the exploration with a sufficiently wide proposal distribution on $\Geff$. We evaluate the convergence of our sampling by demanding $R - 1 \sim \mathcal{O}(10^{-3})$, where $R$ is the Gelman–Rubin diagnostic parameter~\cite{10.1214/ss/1177011136}.

\section{\label{sec:result} Results and Discussion}

As mentioned in Sec.~\ref{sec:model}, for simplicity, our main analyses assume a fixed value of the sum of neutrino masses $\Sigma m_{\nu} = 0.06 \mathrm{eV}$. This in concordance with the fact that current galaxy power spectrum data poorly constrain this parameter \cite{Ivanov:2019hqk,Colas:2019ret}. Nevertheless, for completeness, in App.~\ref{app:mass} we demonstrate that our main results do not depend on this assumption. To avoid clutter, hereafter, we denote the combination of the galaxy power spectrum data $(P_\ell + Q_0 + \mathrm{AP})$ as FS. 

\subsection{\label{sub:profile} Profile likelihood}

\begin{figure*}[htb]
\includegraphics[width=0.975\textwidth]{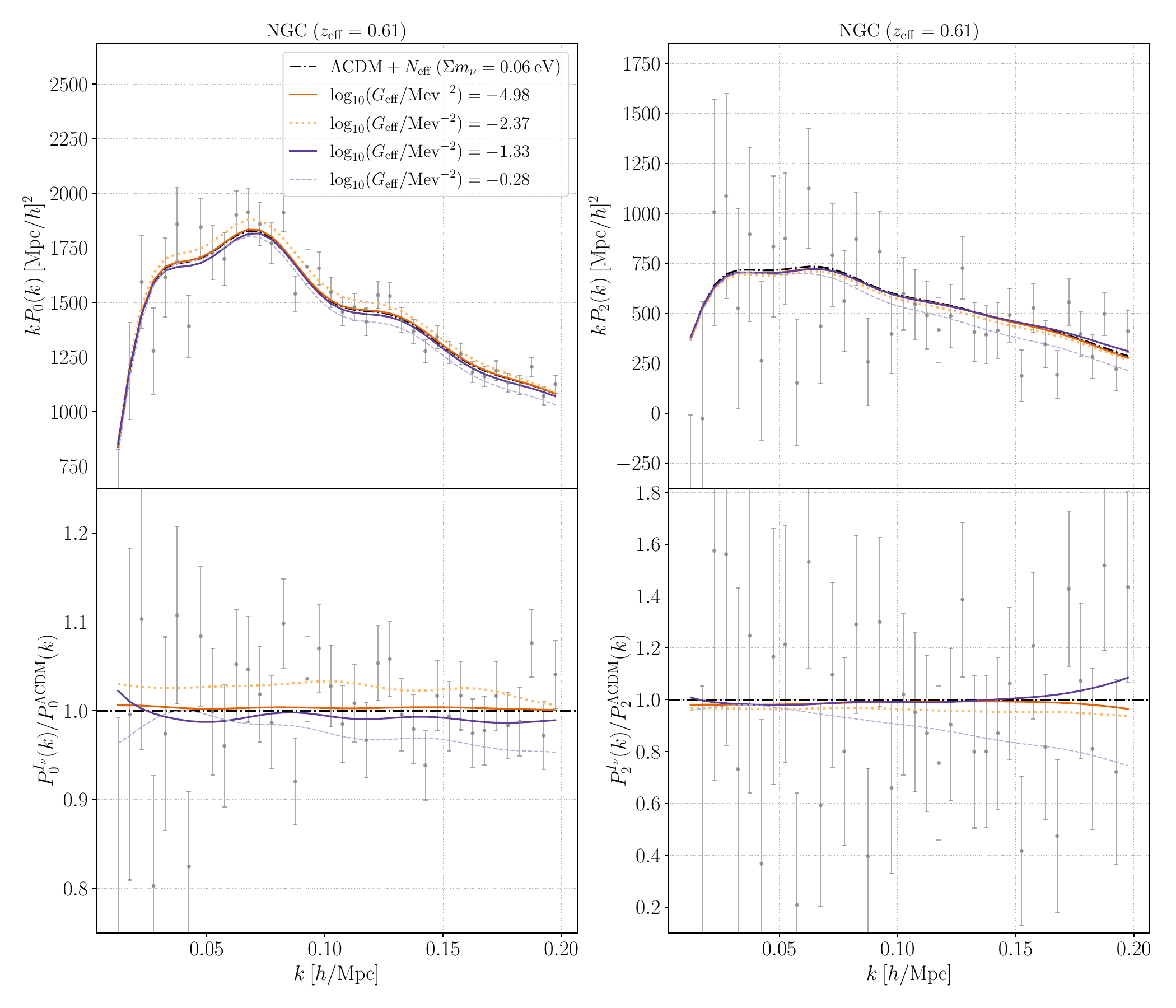}
\caption{\label{fig:multipoles_profile_values_high-z_NGC} Monopole (left) and quadropole (right) of the galaxy power spectrum for some of the $\Delta \chi^2$ extrema obtained through the profile likelihood analysis of FS+BBN data, both for the self-interacting neutrino cosmology and the $\Lambda\mathrm{CDM}+N_\mathrm{eff}$ model. We use $\log_{10}(\Geff/\mathrm{MeV}^{-2}) = \left\lbrace -2.37,-0.28\right\rbrace$ to illustrate regions of the parameter space disfavored by the data, while $\log_{10}(\Geff/\mathrm{MeV}^{-2}) = \left\lbrace -4.98,-1.33\right\rbrace$ to represent the MI$_\nu$ and SI$_\nu$ modes, respectively. The purple solid line shows that strongly self-interacting neutrinos following $\log_{10} (\Geff/\mathrm{MeV}^{-2}) \approx -1.3$ can offer a good fit to the galaxy power spectrum data. The data displayed here correspond to the subset in the NGC at $z_\mathrm{eff} = 0.61$ (see Sec.~\ref{sec:data}).}
\end{figure*}

We profile the likelihood considering both FS data alone and the combination of FS+BBN data. We quantify the goodness-of-fit of the self-interacting neutrino model using  $\Delta \chi^2 \equiv \chi^2_{\mathrm{min},I_{\nu}} - \chi^2_{\mathrm{min},\Lambda\mathrm{CDM}}$. The results of the profile likelihood analyses are presented in Fig.~\ref{fig:profile_likelihood} and Tab.~\ref{tab:fiducial}.

As anticipated in Sec.~\ref{sec:FS}, Fig.~\ref{fig:profile_likelihood} shows that models well-inside the MI$_\nu$ regime provide a fit to the data similar to $\Lambda\mathrm{CDM}$. Indeed, the analysis of the FS+BBN data (blue line and points) shows that models with $\log_{10} (\Geff/\mathrm{MeV}^{-2}) \lesssim -3.5$ lead to a negligible $\Delta \chi^2$, while a slight decrease in $\Delta \chi^2$ can be attained if we solely consider FS data (red line and points). The profile likelihood analysis thus reveals that, when compared with the $\Lambda\mathrm{CDM}+N_\mathrm{eff}$ model, moderately self-interacting neutrino offers at most a marginally better fit to the LSS data.

On the other hand, Fig.~\ref{fig:profile_likelihood} shows that the FS data display a mild preference for the SI$_\nu$ mode. Concretely, we observe that models with $\log_{10}(\Geff/\mathrm{MeV}^{-2}) \approx -1.3$ offer a better fit to the data than $\Lambda\mathrm{CDM}+N_\mathrm{eff}$. The goodness of fit of the SI$_\nu$ mode results in  $\Delta \chi^2= -3.18$ for the analysis of FS data (red line and points) and $\Delta \chi^2= -2.47$ for the analysis of FS + BBN data (blue line and points). Regardless of the constraints imposed by the BBN data, the SI$_\nu$ mode appears to provide a slightly better fit to the galaxy power spectrum data than the standard cosmological model. 

\begin{figure}[htb]
\includegraphics[width=0.995\columnwidth]{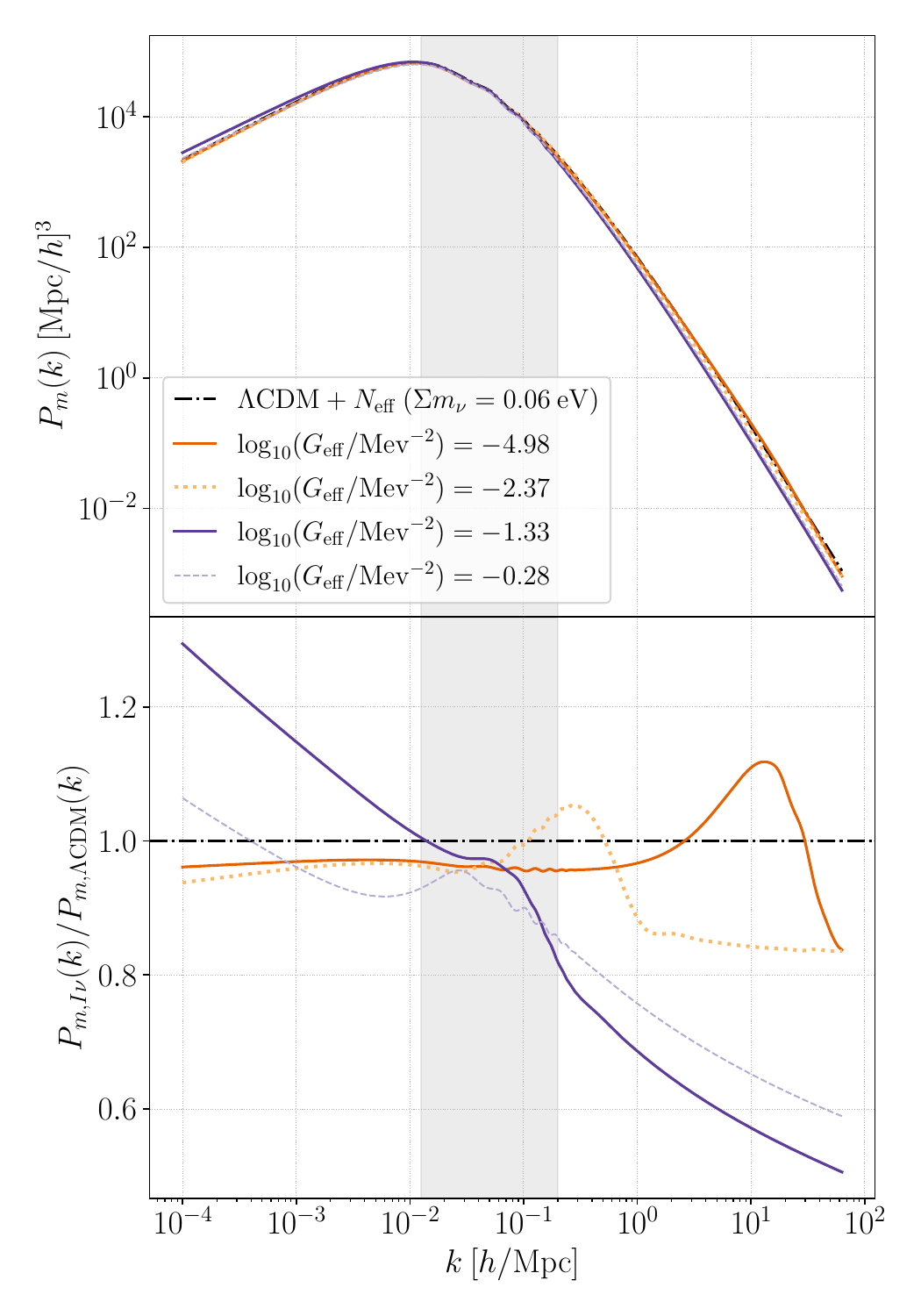}
\caption{\label{fig:Pk_lin_profile_values} Linear matter power spectra for some of the $\Delta \chi^2$ extrema obtained through the profile likelihood analysis of FS+BBN data, both for the self-interacting neutrino cosmology and the $\Lambda\mathrm{CDM}+N_\mathrm{eff}$ model. $\log_{10}(\Geff/\mathrm{MeV}^{-2}) = \left\lbrace -4.98,-1.33\right\rbrace$ are chosen to illustrate the MI$_\nu$ and SI$_\nu$ modes, respectively, while $\log_{10}(\Geff/\mathrm{MeV}^{-2}) = \left\lbrace -2.37,-0.28\right\rbrace$ are used to represent regions of the parameter disfavored by the data. We note that the SI$_\nu$ mode found in the FS+BBN data (dark purple solid line), predicts conspicuous changes in the linear matter power spectrum, including a large suppression of the latter at galactic and sub-galactic scales.}
\end{figure}

To better understand the structure of the likelihood surface as $\Geff$ is varied, we illustrate in Figs.~\ref{fig:multipoles_profile_values_high-z_NGC}~and~\ref{fig:Pk_lin_profile_values} the galaxy and linear matter power spectra, respectively, for some of the $\Delta \chi^2$ extrema obtained through the above profile likelihood analysis. For the better-fitting models, we choose $\log_{10}(\Geff/\mathrm{MeV}^{-2}) = \left\lbrace -4.98,-1.33\right\rbrace$ to illustrate the MI$_\nu$ and SI$_\nu$ modes, respectively. On the other hand, we choose $\log_{10}(\Geff/\mathrm{MeV}^{-2}) = \left\lbrace -2.37,-0.28\right\rbrace$ to represent regions of the parameter space that offer a remarkably worse fit than the standard cosmological model, see Fig.~\ref{fig:profile_likelihood}. The corresponding best-fits to the FS+BBN data for the MI$_\nu$ and SI$_\nu$ modes and the $\Lambda\mathrm{CDM}+N_\mathrm{eff}$ model are displayed in Tab.~\ref{tab:fiducial}. 

In concordance with the discussion presented in Sec.~\ref{sec:FS},  Fig.~\ref{fig:multipoles_profile_values_high-z_NGC} shows that the  MI$_\nu$ and SI$_\nu$ modes (solid dark orange and dark purple lines, respectively) only slightly deviate from the best-fit of the standard cosmological model. In the case of the MI$_\nu$ mode, this behavior is explained by the fact that models with $\log_{10}(\Geff/\mathrm{MeV}^{-2}) \lesssim -3.5$ mostly change the power spectrum at $k \gtrsim 0.5 \: h/\mathrm{Mpc}$, i.e., at (non-linear) scales currently inaccessible by our modeling and observations. In contrast, we observe that the SI$_\nu$ mode attains a good fit to the data by decreasing the values with $A_{\rm s}$ and $n_{\rm s}$, see Tab.~\ref{tab:fiducial}. Notably, this anticorrelation between the SI$_\nu$ regime and the primordial power spectrum parameters, $A_{\rm s}$ and $n_{\rm s}$, has been also observed in the CMB data \cite{Cyr-Racine:2013jua}.

Fig.~\ref{fig:Pk_lin_profile_values} shows that, even when the predictions for the multipoles of the galaxy power spectrum are similar, the underlying linear matter power spectra for the MI$_\nu$ and SI$_\nu$ modes are significantly different. Indeed, owing to the decrease in $A_{\rm s}$ and $n_{\rm s}$, the SI$_\nu$ mode predicts a $\gtrsim 30\%$ ($\gtrsim 40\%$) suppression of the power spectrum at galactic (sub-galactic) scales while exhibiting a barely visible bump that peaks around $k \approx 0.1 \: h/\mathrm{Mpc}$. This model also displays an increase in power at very large scales. Remarkably, this general structure of the SI$_\nu$ matter power spectrum matches that found in Ref.~\cite{Kreisch:2019yzn} using CMB data only.  On the other hand, the MI$_\nu$ mode features a bump that peaks well inside the non-linear scales and a modest and nearly constant suppression of the power spectrum for modes~$k \lesssim 1 \: h/\mathrm{Mpc}$.

\subsection{\label{sub:constraints} Cosmological constraints}

We scan the parameter space of the self-interacting neutrinos and $\Lambda\mathrm{CDM}+N_\mathrm{eff}$ models using the Metropolis-Hasting algorithm implemented in \texttt{montepython}~\cite{Audren:2012wb,Brinckmann:2018cvx}. We perform the exploration imposing a flat prior in the self-coupling strength $\log_{10} (\Geff) = [-5.5,0.5]$ and the effective number of relativistic species $N_\mathrm{eff} = [2.013,5.513]$. The other cosmological parameters are set to follow improper flat priors. Furthermore, we impose priors on the nuisance parameters of the BOSS likelihood following Ref.~\cite{Philcox:2021kcw}. To illustrate the role of each data set in constraining the delay in the free streaming of neutrinos, we perform several analyses considering different combinations of the data. Our results are shown in Figs.~\ref{fig:square_nu_all_cases},~\ref{fig:LSS_vs_CMB_v2}~and~\ref{fig:triangle_tensions_nu_int_vs_LCDM} and in Tab.~\ref{tab:constraints}. Unless otherwise stated, we conservatively assume $k_\mathrm{max} = 0.20 \: h/\mathrm{Mpc}$.

\subsubsection{The role of the linear and (mildly) non-linear scales}

\begin{figure*}[htb]
\includegraphics[width=0.975\textwidth]{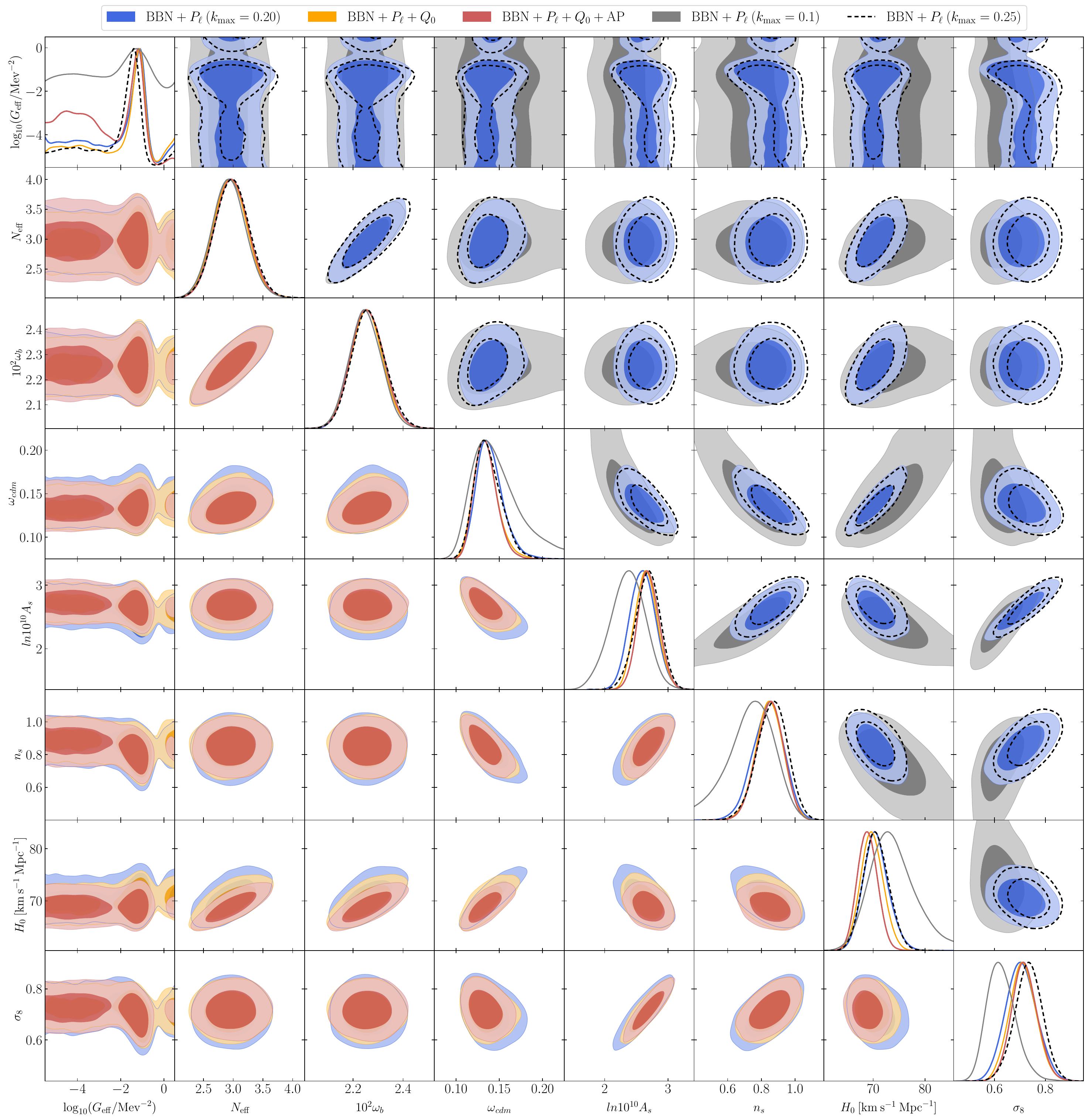}
\caption{\label{fig:square_nu_all_cases} Marginalized constraints, at $68\%$ and $95\%$ confidence levels, on the cosmological parameters of the self-interacting neutrino model when considering different combinations of data. The upper triangular portion highlights the role of the linear and mildly non-linear scales in the task of constraining a delay in the onset of the free streaming of neutrinos, while the lower triangular portion emphasizes the contribution of the $Q_0$ and AP data.}
\end{figure*}

Panels in the upper triangular portion of Fig.~\ref{fig:square_nu_all_cases} show the constraints obtained from the analysis of BBN + $P_\ell$ data when different values for $k_\mathrm{max}$ are assumed. We note that data merely considering modes belonging to the linear scale, i.e. $k_\mathrm{max} = 0.1 \: h/\mathrm{Mpc}$, 
do not constrain the self-coupling constant $\Geff$ (gray contours and lines). However, the situation significantly changes if we include modes associated with the mildly non-linear scales, that is, if we adopt $k_\mathrm{max} = 0.20 \: h/\mathrm{Mpc}$ (blue contours and lines). In such a case, we not only observe an improvement in the constraints of all the cosmological parameters in general but also a net decrease of the posterior for models belonging to the MI$_\nu$ regime. Marginal improvements to the latter result are obtained if we consider $k_\mathrm{max} = 0.25 \: h/\mathrm{Mpc}$ (dashed black contours and lines). It is important to note that, regardless of the value of $k_\mathrm{max}$, we discern the existence of a non-trivial correlation between $A_{\rm s}$ and $n_{\rm s}$ and the strongly interacting neutrinos.

\begin{table*}[htb]
\caption{\label{tab:constraints}
68$\%$ confidence level intervals for the cosmological parameters obtained from the analysis of the FS+BBN data for the different cosmologies here considered. Constraints on the mild and strong regimes are obtained by splitting the sampling into two subsets, one following $ \log_{10}(\Geff/\mathrm{MeV}^{-2}) \leq -2.5$ and the other the opposite, respectively. We highlight that SI$_\nu$ cosmologies lead to lower values of $A_{\rm s}$ and $n_{\rm s}$.}
\begin{ruledtabular}
\begingroup
\renewcommand{\arraystretch}{1.35}
\begin{tabular}{lccc}
\hline
Parameter & $\Lambda\mathrm{CDM}+N_\mathrm{eff}$ & MI$_\nu$ regime & SI$_\nu$ regime \\
\hline
$\log_{10}(G_\mathrm{eff}/\mathrm{Mev}^{-2})$ & - & $-4.07^{+0.77}_{-1.1}$ & $-1.30^{+0.47}_{-0.37}$ \\
$10^{2}\omega_{b}$ & $2.259\pm 0.063$ & $2.256\pm 0.065$ & $2.257\pm 0.064$ \\
$\omega{}_{cdm}$ & $0.134^{+0.011}_{-0.014}$ &  $0.134^{+0.011}_{-0.015}$ & $0.135^{+0.010}_{-0.014}$ \\
$\ln \left(10^{10}A_{s}\right)$ & $2.73\pm 0.16$ & $2.73^{+0.15}_{-0.18}$ & $2.64\pm 0.16$ \\
$n_{s}$ & $0.882\pm 0.069$ & $0.881\pm 0.071$ & $0.813\pm 0.072$ \\
$H_0\; [\mathrm{km}\;\mathrm{s}^{-1}\;\mathrm{Mpc}^{-1}]$ & $68.9^{+1.8}_{-2.0}$ & $68.9\pm 1.9$ & $69.0\pm 1.9$ \\
$N_\mathrm{eff}$ & $2.98^{+0.25}_{-0.29}$ & $2.97\pm 0.28$ & $2.98\pm 0.27$ \\
$\sigma_8$ & $0.725^{+0.044}_{-0.050}$ & $0.730^{+0.042}_{-0.053}$ & $0.702\pm 0.051$ \\
\hline
\end{tabular}
\endgroup
\end{ruledtabular}
\end{table*}

\subsubsection{The effects of the $Q_0$ estimator and the AP test}

We show in the panels of the lower triangular portion of Fig.~\ref{fig:square_nu_all_cases} the results of the analyses when considering the $Q_0$ and AP data. We observe that inclusion of the $Q_0$ estimator data (light orange contours and lines) marginally improves the results of the $\mathrm{BBN} + P_{\ell}$ analysis (blue contours and lines). This is not surprising since although the $Q_0$ estimator allows us to probe smaller scales, corresponding to wavenumber up to $k = 0.4 \: h/\mathrm{Mpc}$, its current BOSS DR12-based estimative is shot noise-dominated~\cite{Ivanov:2021fbu}. 

On the other hand, the inclusion of the AP data (red contours and lines) increases the probability density distribution of the self-coupling constant around the MI$_\nu$ regime and slightly decreases the likelihood of models with $\log_{10}(\Geff/\mathrm{MeV}^{-2}) \gtrsim -0.5$. Nonetheless, the AP data do not disfavor the SI$_\nu$ mode, which still offers a slightly better fit to the data than the MI$_\nu$ mode.

We present the constraints obtained from the analysis of $\mathrm{BBN}+P_\ell +Q_0+\mathrm{AP}$ data for both the $\Lambda\mathrm{CDM}+N_\mathrm{eff}$ and $I_\nu$ models in Tab.~\ref{tab:constraints}. Additionally, we provide the constraints derived for the MI$_\nu$ and SI$_\nu$ modes. Such constraints are obtained by splitting the sampling into two subsets, one corresponding to $ \log_{10}(\Geff/\mathrm{MeV}^{-2}) \leq -2.5$ and the other to the opposite. We argue that this mode separation scheme is enough to provide an insight into the parameter space of the mild and strong interacting neutrino cosmologies. As in the case of the profile likelihood analysis, the fourth column in Tab.~\ref{tab:constraints} shows the SI$_\nu$ mode leads to lower values of the primordial power spectrum parameters: $A_{\rm s}$ and $n_{\rm s}$. 

\begin{figure}[t!]
\includegraphics[width=0.95\columnwidth]{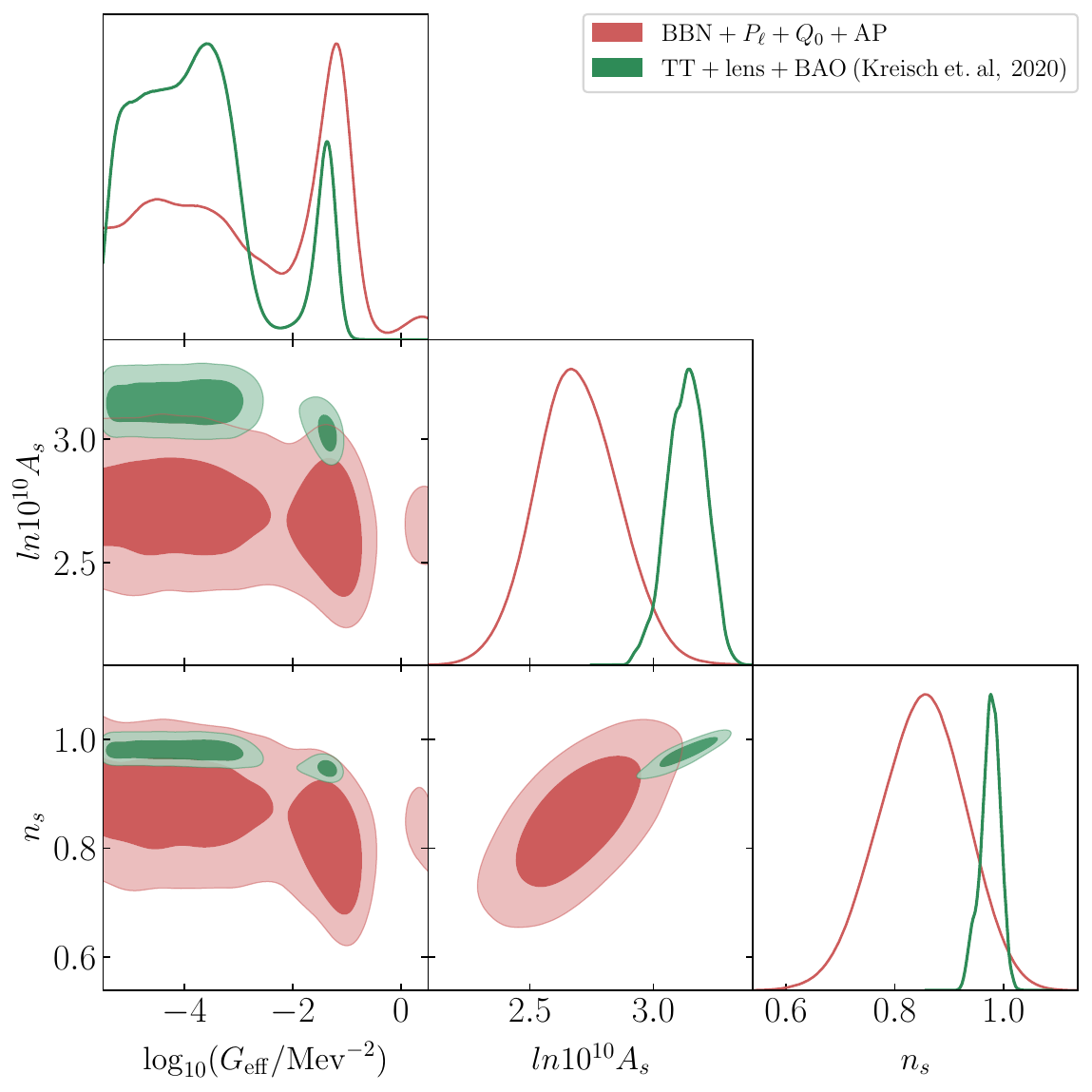}
\caption{\label{fig:LSS_vs_CMB_v2} Comparison between the marginalized constraints, at $68\%$ and $95\%$ confidence levels, on different parameters of the $I_\nu$ model obtained from our main analysis and one of the analyses presented in Ref.~\cite{Kreisch:2019yzn}.}
\end{figure}

We emphasize that our results and previous analyses of CMB data not only reveal the presence of the SI$_\nu$ but also agree that there exists an anticorrelation among $\Geff$ and the amplitude and tilt of the primordial power spectrum, $A_s$ and $n_s$, respectively. We explicitly illustrate this in Fig.~\ref{fig:LSS_vs_CMB_v2}, where we compare the underlying constraints obtained from the analysis of BBN + $P_{\ell}$ + $Q_0$ + AP data with the ones obtained in Ref.~\cite{Kreisch:2019yzn} from the analysis of the CMB TT + lens + BAO data. This comparison hints that it is possible to create a self-consistent picture for strongly self-interacting neutrinos, implying that cosmological data could allow a cosmological scenario in which neutrino free streaming is delayed until close to the matter-radiation equality epoch.

\subsubsection{Pondering the cosmological tensions}

Finally, we assess the matter of cosmological tensions. As stated before, owing to correlations with $A_{\rm s}$ and $n_{\rm s}$, the SI$_\nu$ mode leads to a power spectrum that is significantly suppressed at small scales, see Fig.~\ref{fig:Pk_lin_profile_values}. Tab.~\ref{tab:constraints} shows that when compared to the $\Lambda\mathrm{CDM}+N_\mathrm{eff}$ cosmology, the latter produces a $\sim 3\%$ decrease in $\sigma_8$, the root mean square of the matter fluctuations at $8\:h/\mathrm{Mpc}$. Table~\ref{tab:constraints} also shows that, regardless of the model, BOSS data consistently yield a lower value of $\sigma_8$ when compared to the CMB constraints.

\begin{figure*}[t!]
\includegraphics[width=0.975\textwidth]{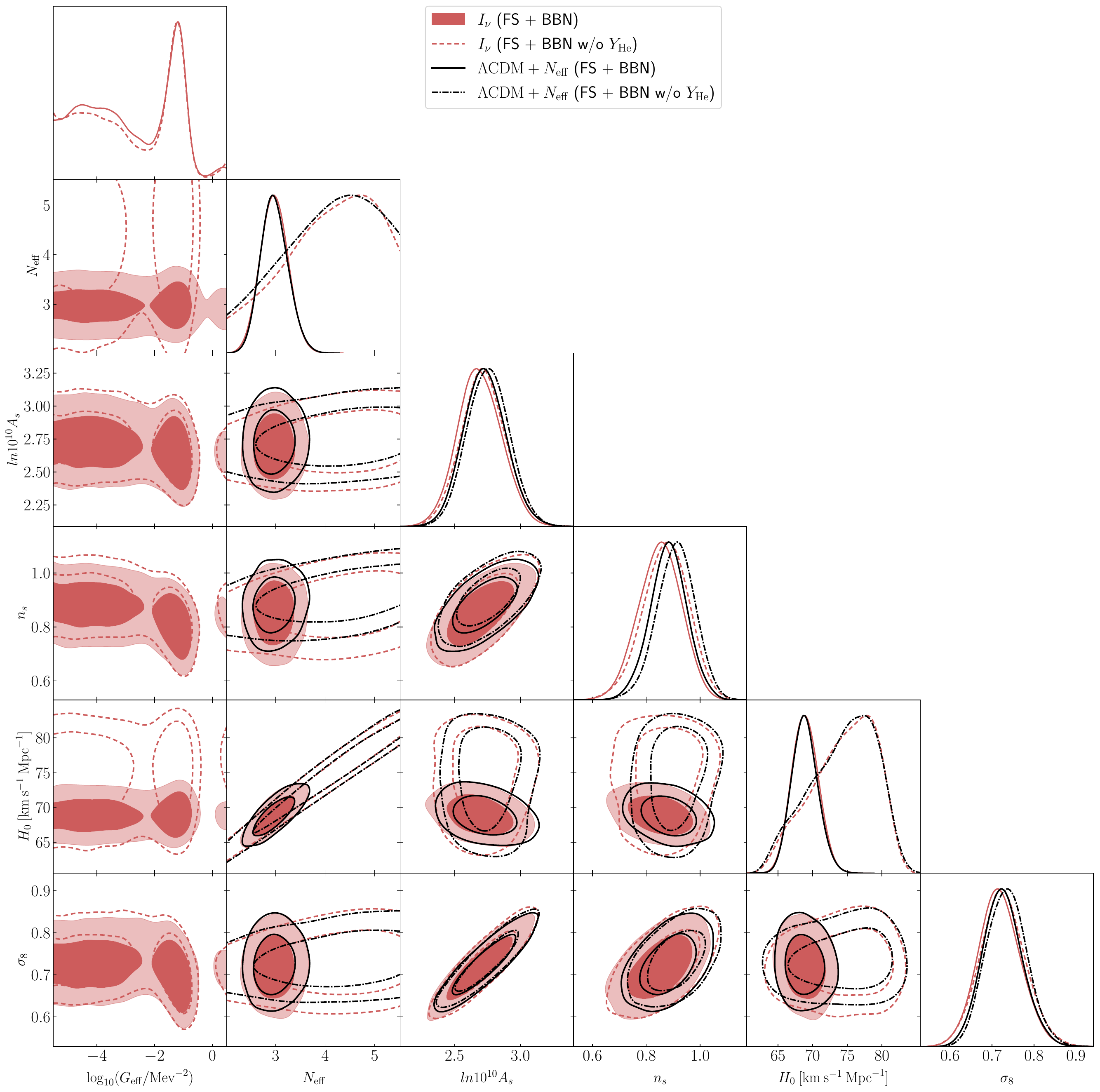}
\caption{\label{fig:triangle_tensions_nu_int_vs_LCDM} Marginalized constraints, at $68\%$ and $95\%$ confidence levels, on select parameters of the  $\Lambda\mathrm{CDM} + N_\mathrm{eff}$ and $I_\nu$ models obtained from the analysis of FS+BBN data with and without the inclusion of the primordial helium abundance,~$Y_\mathrm{He}$.}
\end{figure*}

Furthermore, from Tab.~\ref{tab:constraints}, one can note that the $\Lambda\mathrm{CDM}+N_\mathrm{eff}$ model and the self-interacting neutrino cosmologies produce indistinguishable values for $H_0$. This similarity arises because, in both scenarios, the $N_\mathrm{eff}$ value is tightly constrained by the primordial abundance of helium $Y_\mathrm{He}$, resulting in nearly identical sizes of the baryon-photon sound horizon; the change induced by $\Geff$ in the SI$_\nu$ case is subdominant. To better illustrate this point, we perform an extra analysis when considering BBN observations without the presence of $Y_\mathrm{He}$ data. The results of this analysis are shown in Fig.~\ref{fig:triangle_tensions_nu_int_vs_LCDM}. 

Fig.~\ref{fig:triangle_tensions_nu_int_vs_LCDM} offers a direct comparison between the $I_\nu$ and the $\Lambda\mathrm{CDM} + N_\mathrm{eff}$ models when considering BBN data with (solid lines and contours) and without (dashed lines and contours) observations of $Y_\mathrm{He}$.  We immediately observed that removing the helium abundance constraint frees the values of $N_\mathrm{eff}$ and $H_0$ in both models, leaving them largely unconstrained by the FS data. This is the result of a well-known geometric degeneracy between the baryon-photon sound horizon (which can be adjusted by changing $N_\mathrm{eff}$) and the angular diameter distance (which scales as $H_0^{-1}$) \cite{Aylor:2018drw}. This highlights the importance of BBN \cite{Cyr-Racine:2021oal}, and more generally, of our assumptions about the physics of the early Universe, to the value of the Hubble constant inferred from FS data. Thus, FS data by themselves cannot weigh in on whether interacting neutrinos may play a role in the current discrepancy between different measurements of the Hubble constant. We note however that the posterior distribution of $\Geff$ is nearly independent of whether a BBN prior is assumed or not. 

\section{\label{sec:conclu} Conclusions}

Several analyses have pointed out that some cosmological data show a preference for a cosmological scenario in which the free streaming of neutrinos is delayed until close to the epoch of matter-radiation equality. Produced by yet-unknown strong self-interactions in the neutrino sector, this nonstandard scenario generates important changes in the evolution of cosmological perturbations at linear and nonlinear scales that could impact the LSS of the Universe. Here, we have investigated if LSS data are sensitive to these changes. We adopted the simplest cosmological representation for self-interacting neutrinos and later analyzed the Full Shape of the galaxy power spectrum, and BBN data, within this context. Remarkably, our analysis unveils the presence of the SI$_\nu$ mode in the galaxy power spectrum data and adds a new chapter to the tale of the two modes.

Indeed, we have found that self-interacting neutrinos with $\log_{10}(\Geff/\mathrm{MeV}^{-2}) \approx - 1.3$, provide a good fit to the galaxy power spectrum data, regardless of the presence or absence of BBN priors. The goodness of fit of such a scenario has been quantified to be $\Delta \chi^2 \approx -2.5$ ($\Delta \chi^2 \approx -3$) when BBN priors are (not) taken into account in the analysis, thus displaying a modest preference for strongly interacting scenarios over the $\Lambda\mathrm{CDM}+N_\mathrm{eff}$ and MI$_\nu$ models. Moreover, we have exposed that this modest preference for the SI$_\nu$ is driven by the data in the mildly nonlinear scales.  

Compared to the $\Lambda\mathrm{CDM}+N_\mathrm{eff}$ model, the SI$_\nu$ mode found in the galaxy power spectrum data displays a significant matter clustering suppression on small scales. Such a suppression, that is shown to be greater than $40\%$ at sub-galactic scales, i.e, $k \gtrsim 10\: h/\mathrm{Mpc}$, is driven by the underlying decrease of $A_{\rm s}$ and $n_{\rm s}$ that strongly self-interacting neutrinos prefer. Importantly, the same predicted suppression of small-scale power also appears in previous analyses that rely on the CMB observations~\cite[see Ref.][for instance]{Kreisch:2019yzn}. This lack of small-scale power, although not as dramatic as in e.g.~warm dark matter models \cite{Bond_1983,Bode:2000gq,Dalcanton:2000hn}, could be probed via substructure lensing (see e.g.~Refs.~\cite{vegetti2012,vegetti2014,Gilman:2019nap,Hsueh:2019ynk,Zhang:2022djp,Sengul:2022edu,Wagner-Carena_2023}) or observations of the Milky Way satellites (see e.g.~Refs.~\cite{Newton_2018,Nadler:2018iux,Nadler:2019zrb,Nadler_2021}) . 

We conclude that our results, which are consistent across both profile likelihood and Bayesian exploration analyses, do not only expose the presence of the persistent SI$_\nu$ mode in the galaxy power spectrum data but also suggest that cosmological data can potentially accommodate a self-consistent cosmological scenario in which the onset of the free streaming of neutrinos is delayed until close to the matter-radiation equality epoch. Although this finding does not pose an immediate issue for the $\Lambda$CDM model (the statistical preference being mild), our analysis deepens the riddle around the two-mode puzzle as we now have two different kinds of cosmological data (CMB and galaxy clustering) showing some preference for the SI$_\nu$. In line with this, we would like to bring back attention to one of the conclusions presented by Ref.~\cite{Kreisch:2019yzn}: while we typically explore new physics by proposing mild deformations of the $\Lambda$CDM model, it is crucial to bear in mind that radically different scenarios could provide a good fit to the cosmological observables. Thus, our results motivate the thorough exploration of neutrino interaction models capable of reconciling all CMB and LSS data in the SI$_\nu$ regime, including CMB polarization data from Planck \cite{Planck:2018vyg}. Since polarization data are particularly sensitive to the anisotropic stress history of the Universe, they naturally are better probes of the flavor structure of the neutrino interactions. The fact that such data disfavor the simplest universal model considered here indicates that a more complex (and realistic) neutrino interaction model that includes a strong flavor dependence might be preferred. We leave to future work the study of a model capable of accommodating all CMB and LSS data, while not running afoul of other laboratory constraints.

It is also interesting to comment on how our results connect to previous free-streaming phase shift analyses showing consistency between the SM predictions and both CMB \cite{Follin:2015hya} and BAO \cite{Baumann:2017lmt,Baumann:2019keh} data. These works use a one-parameter family of templates calibrated to $\Lambda$CDM to measure the neutrino-induced phase shift from the data, phrasing their results in terms of the effective number of neutrino species, $N_{\rm eff}$. By construction, such templates can only capture scenarios in which the free-streaming radiation fraction is constant in the era after BBN but prior to the epoch of recombination. The time-varying free-streaming fraction caused by the late neutrino decoupling we studied here leads to a phase shift structure of the CMB and BAO peaks that is distinct from that captured by the templates used so far, leaving them unable to directly capture the SI$_\nu$ signal. In principle, phase-shift templates capable of capturing this time-dependent free-streaming fraction could be built, and the possible presence of the SI$_\nu$ could be studied by isolating its impact on the phase of CMB and BAO peaks. We leave such an analysis to future works.

Finally, now that we have established the existence of the SI$_\nu$ mode in two independent cosmological data sets, we can discard the possibility that its existence is caused by an accidental feature in the CMB sky. The apparent consistency between some CMB data and the large-scale distribution of galaxies indicates that the SI$_\nu$, whatever its microscopic origin is, is an actual physical feature present in the data. While we have explored this feature here using the language of self-interacting neutrinos, it is also possible that our results are hinting at the existence of a yet-to-be-discovered early-Universe phenomenon that is not related at all to new physics in the neutrino sector. Our results highlight the need for considering a broader range of phenomenologies deep in the radiation-dominated epoch that could be consistent with current cosmological observations. 

\begin{acknowledgments}
We thank Vera Gluscevic, Adam He, and Daniel Green for useful comments on an initial version of this manuscript. This work was supported by the National Science Foundation (NSF) under grant AST-2008696 and the REU site grant PHY-1659618. D.~C.~and F.-Y.~C.-R. would also like to thank the Robert E.~Young Origins of the Universe Chair fund for its generous support. We also would like to thank the UNM Center for Advanced Research Computing, supported in part by the NSF, for providing the research computing resources used in this work.\\
\end{acknowledgments}

\appendix

\section{\label{app:mass} The impact of Neutrino mass}

\begin{figure*}[htb]
\includegraphics[width=0.85\textwidth]{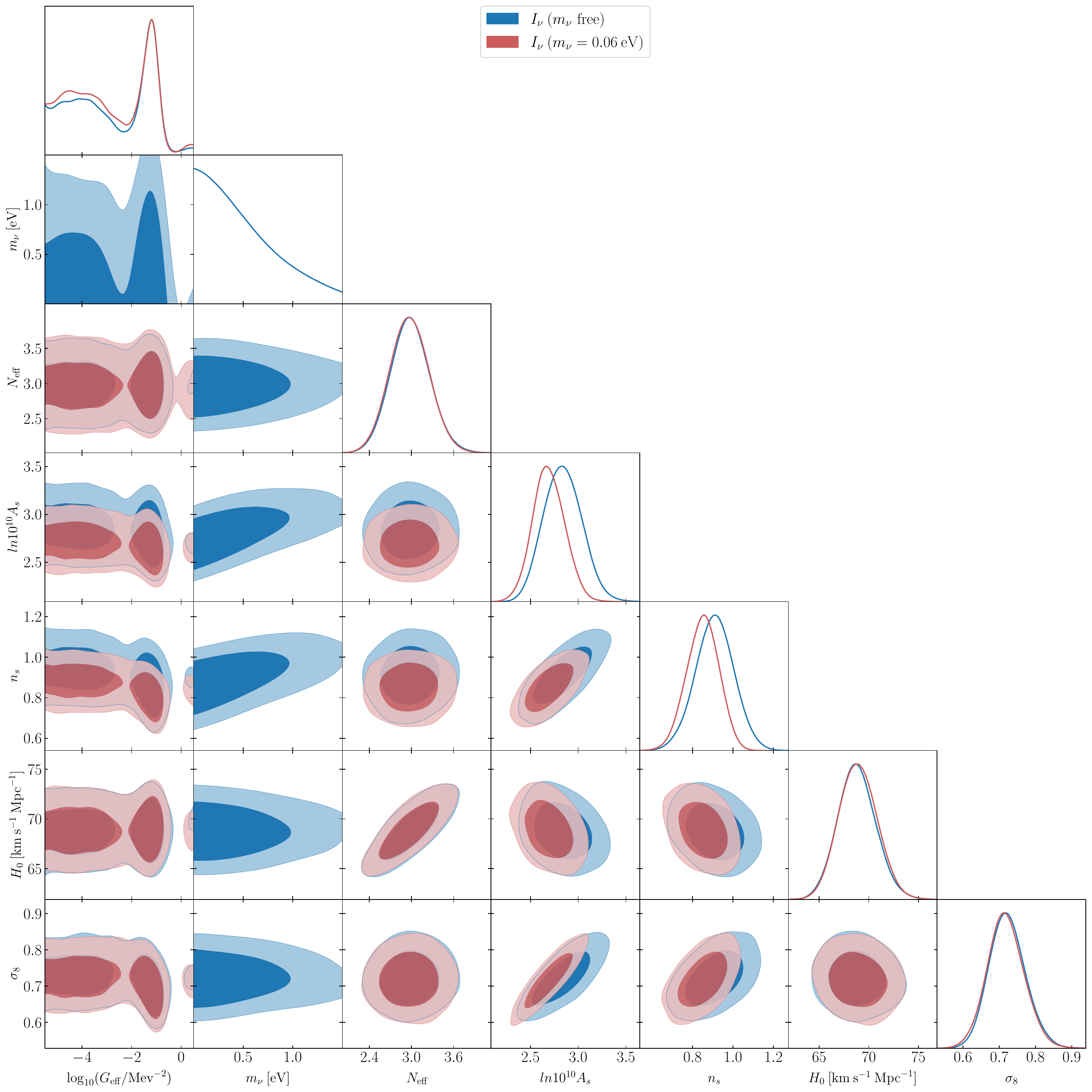}
\caption{\label{fig:triangle_nu_mass_assumption} Marginalized constraints, at $68\%$ and $95\%$ confidence level, on selected cosmological parameters of the $I_\nu$ model obtained from the analysis of the FS+BBN data when assuming the total neutrino mass as fixed $\Sigma_{\nu} = 0.06$ eV and as a free parameter.}
\end{figure*}

Our main results rely on the assumption of a fixed value for the sum of neutrino masses, more precisely, $\Sigma m_\nu = 0.06$ eV. However, to illustrate that this assumption does not bias our results and conclusions, we have carried an extra analysis of the FS+BBN data assuming $\Sigma m_\nu$ as free parameter. The results of this complementary analysis is shown in Fig.~\ref{fig:triangle_nu_mass_assumption}. 

We note that the assumption of a fixed value for the sum of neutrino masses does not significantly affect the constraints in $\Geff$, $N_\mathrm{eff}$ or the cosmological derived parameters of interested $H_0$ and $\sigma_8$. Nonetheless, this assumption leads to slightly smaller values of the amplitude, $A_{\rm s}$, and tilt, $n_{\rm s}$, of the primordial power spectrum.

\bibliography{apssamp}

\end{document}